\documentclass[review]{elsarticle}
\usepackage{enumerate}
\usepackage{booktabs}
\usepackage{amsxtra}
\begin{document}
\begin{frontmatter}
%%%%%%%%%%%%%%%%%%%%%%%%TITLE%%%%%%%%%%%%%%%%%%%%%%%%
\title{Turbulent Transport in Tokamak-plasmas: \\ A Thermodynamic Approach}
%%%%%%%%%%%%%%%%%%%%%%END_TITLE%%%%%%%%%%%%%%%%%%%%%%%

%%%%%%%%%%%%%%%%%%%%%%ADDRESSES%%%%%%%%%%%%%%%%%%%%%%
\author{Giorgio SONNINO\corref{cor1}}
\cortext[cor1]{Corresponding author}
\ead{giorgio.sonnino@ulb.be}
\author{Philippe PEETERS\corref{}}
\ead{peeters.philippe@gmail.com}
\author{Pasquale NARDONE\corref{}}
\ead{pasquale.nardone@ulb.}
\author{\\ and Enrique TIRAPEGUI$^{\dag}$}
\address{
Universit{\'e} Libre de Bruxelles (U.L.B.)\\
Facult{\'e} de Sciences, Department of Physics 
\\Campus de la Plaine C.P. 231 - Bvd du Triomphe
B-1050 Brussels - Belgium\\
\vskip1truecm
\noindent Manuscript Accepted for Publication in the review Elsevier - Chaos, Solitons $\&$ Fractals}

%%%%%%%%%%%%%%%%%%%%END_ADDRESSES%%%%%%%%%%%%%%%%%%%%%
%%%%%%%%%%%%%%%%%%%%%%ABSTRACT%%%%%%%%%%%%%%%%%%%%%%%

\begin{abstract}
In previous work we provided the explicit form of the nonlinear PDEs, subjected to the appropriate boundary conditions, which have to be satisfied by transport coefficients for systems out of Onsager's region. Since the proposed PDEs are obtained without neglecting any term present in the balance equations (i.e., the mass, momentum, and energy balance equations), we propose them as a good candidate for describing also transport in thermodynamic systems in turbulent regimes. As a special case, we derive the nonlinear PDEs for transport coefficients when the thermodynamic system is subjected to two thermodynamic forces. In this case, the obtained PDE is, in thermodynamical field theory (TFT), analogous to Liouville's equation in Riemannian (or pseudo-Riemannian) geometry. The validity of our model is tested by analyzing a concrete example where Onsager's relations manifestly disagree with experience: transport in Tokamak-plasmas. More specifically, we compute the electron mass and energy losses in turbulent FTU (\textit{Frascati Tokamak Upgrade})-plasmas. We show the agreement between the theoretical predictions and experimental observations. This approach allows for predicting the values of the Bohm and the gyro-Bohm coefficients. To the best of our knowledge, it is the first time that such coefficients have been evaluated analytically. The aim of this series of works is to apply our approach to the {\it Divertor Tokamak Test facility} (DTT), to be built in Italy, and to ITER.
\end{abstract}

\end{frontmatter}

%%%%%%%%%%%%%%%%%%%%%END_ABSTRACT%%%%%%%%%%%%%%%%%%%%%

%%%%%%%%%%%%%%%%%%%%%%TEXT_PAPER%%%%%%%%%%%%%%%%%%%%%%
\section{Introduction}\label{I}
The first attempts to develop non-equilibrium thermodynamics theory occurred after the first observations of some coupled phenomena of thermal diffusion and thermoelectric. However, the major obstacle to overcome is that the number of unknowns is greater than the number of equations expressing the conservation laws. When there are more unknowns than equations expressing conservation laws supplementary \textit{closure relations} are needed to make the problem solvable. Generally, these additional closure laws are not derivable from one of the physical equations being solved. Several approaches to getting closure relations are currently applied. Among them we cite the so-called \textit{troncation schemes} and the \textit{Asymptotic schemes}. In \textit{truncation schemes}, higher order moments are arbitrarily assumed to vanish, or simply negligible with respect to the terms of lower moments. Truncation schemes can often provide quick insight into fluid systems but always involve uncontrolled approximation. This method is often used in transport processes in Tokamak-plasmas (see, for instance, the book \cite{balescu2}). The \textit{asymptotic schemes} is based on the rigorous exploitation of some small parameters. They have the advantage of being systematic and providing some estimate of the error involved in the closure. However, as the title itself suggests, these methods are effective only when small parameters enter, by playing a crucial role, in the dynamic equations. These schemes are often used for solving numerically kinetic equations (ref., for instance, to the book \cite{carillo}). Another possibility is to obtain closure relations by formulating a specific theory or \textit{ad hoc} models. The most important closure equations are the so-called \textit{transport equations} (or the \textit{flux-force relations}), relating the thermodynamic forces with the conjugate dissipative fluxes that produce them. The thermodynamic forces are related to the spatial inhomogeneity and (in general) they are expressed as gradients of the thermodynamic quantities. The study of these relations is the object of \textit{non-equilibrium thermodynamics}. Morita and Hiroike eased this task for a closure relation by providing the formally exact closure formula \cite{morita}
\begin{equation}\label{I1}
J_\nu(X)=\tau_{\mu\nu}(X)X^\mu
\end{equation}
\noindent Here, $X^\mu$ and $J_\mu$ denote the thermodynamic forces and thermodynamic fluxes, respectively. Coefficients $\tau_{\mu\nu}(X)$ are the \textit{transport coefficients}, where it is clearly highlighted that the transport coefficients may depend on the thermodynamic forces. We suppose that all quantities appearing in Eq.~(\ref{I1}) are dimensionless. Besides, in this equation, as well as in the sequel, the Einstein summation convention on the repeated indices is understood. Matrix $\tau_{\mu\nu}(X)$ can be decomposed in two pieces, one symmetric and the other skew-symmetric, which we denote with $g_{\mu\nu}(X)$ and $f_{\mu\nu}(X)$, respectively. The second law of thermodynamics requires that $g_{\mu\nu}(X)$ is a positive-definite matrix. Note that, in general, the dimensionless entropy production, denoted by $\sigma$, with $\sigma=\tau_{\mu\nu}(X) X^\mu X^\nu =g_{\mu\nu}(X)X^\mu X^\nu$, may not be a simply bilinear expression of the thermodynamic forces (since the transport coefficients may depend on the thermodynamic forces). For conciseness, in the sequel we drop the symbol $X$ in $g_{\mu\nu}$ as well as in the skew-symmetric piece of the transport coefficients $f_{\mu\nu}$ being implicitly understood that these matrices may depend on the thermodynamic forces. Close to equilibrium, the transport equations of a thermodynamic system are provided by the well-known Onsager theory:
\begin{equation}\label{I2}
J_\mu=\tau_{0\mu\nu}X^\nu
\end{equation}
\noindent where $\tau_{0\mu\nu}$, that denotes the transport coefficients in Onsager's region, {\it does not depend on the thermodynamic forces}. Of course, even matrix $\tau_{0\mu\nu}$ can be decomposed into a sum of two matrices, one symmetric and the other skew-symmetric, which we denote with $L_{\mu\nu}$ and $f_{0\mu\nu}$, respectively. The second law of thermodynamics requires that $L_{\mu\nu}$ is a positive definite matrix. The region where Eq.~(\ref{I2}) hold, where $\tau_{0\mu\nu}$ does not depend on the thermodynamic forces, is called {\it Onsager's region} or, the {\it linear region of thermodynamics} \cite{onsager1}, \cite{onsager2}.

\subsection{Why study Tokamak-plasmas?}

\begin{figure}[h]
%\sidecaption
\centering\includegraphics[scale=.20]{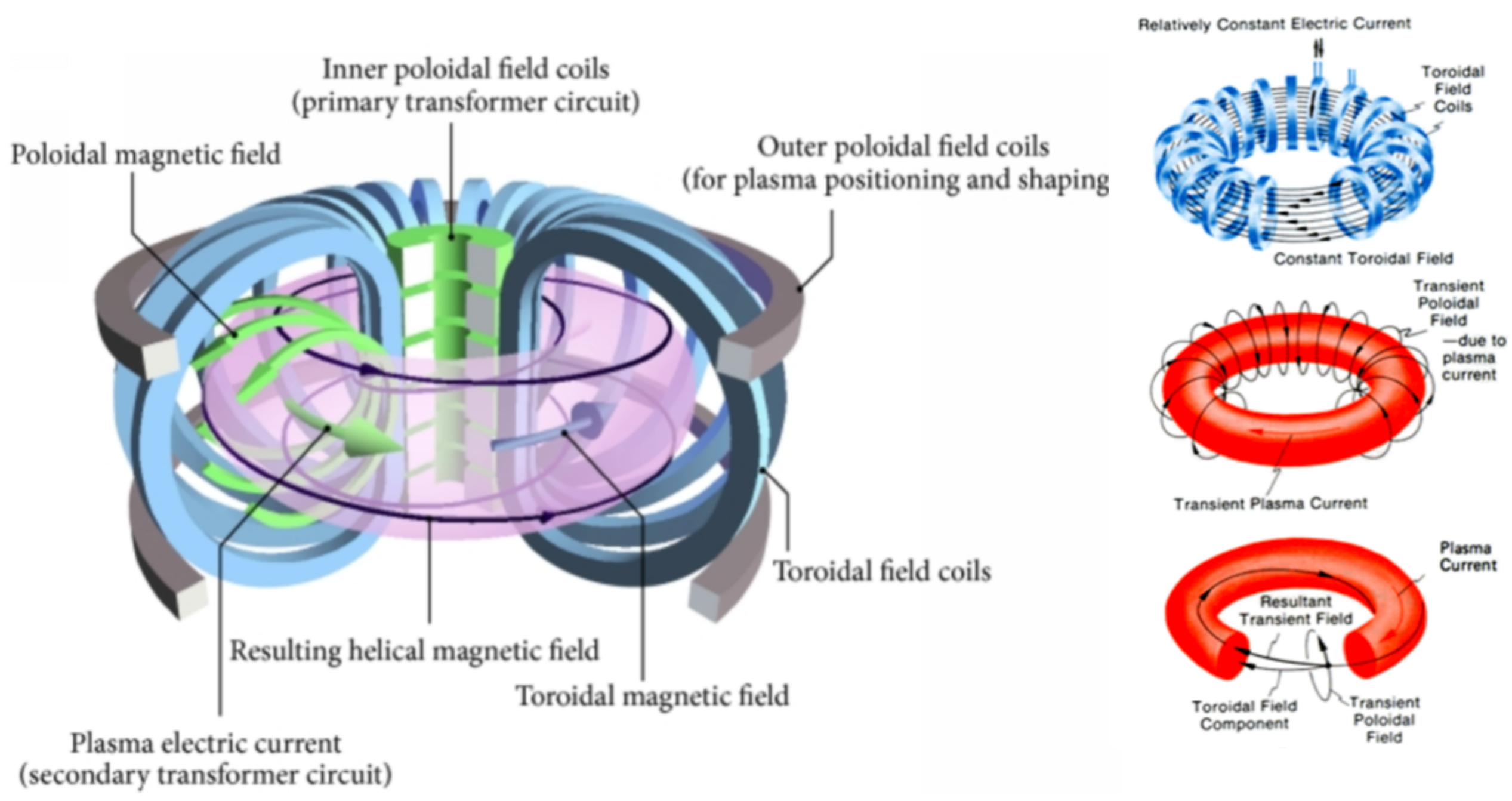}
\caption{{\bf Schematic representation of a Tokamak.} The figure shows the toroidal and the poloidal coils and the toroidal and poloidal magnetic-field lines (black) in a fusion device (image from https://da.wikipedia.org/wiki/Tokamak).}
\label{fig_tokamak}
\end{figure}

\noindent When heated to fusion temperatures, the electrons in atoms disassociate, resulting in a fluid of nuclei and electrons known as a plasma. Unlike electrically neutral atoms, a plasma is electrically conductive, and can, therefore, be manipulated by electric and/or magnetic fields. Magnetic confinement fusion devices exploit the fact that charged particles in a magnetic field experience a Lorentz force and follow helical paths along the field lines. A Tokamak is a device which uses a powerful magnetic field to confine plasma in the shape of a torus (see Fig.~\ref{fig_tokamak}).
%%%%%%%%%%%%%%%%%%%%%%%%%%%%%%%%%%%%%%%%%%
%%%%%%%%%%%%%%%%%%%%%%%%%%%%%%%%%%%%%%%%%%
\noindent In Tokamak-plasma, collisions tend to bring the system towards the local equilibrium while free flow and magnetic field in a curved configuration tend to push the system far from equilibrium. The final result is that plasma is far from equilibrium. More specifically, {\it magnetically confined plasmas in fusion devices are typical examples of systems out of Onsager's region}. Since, the transport coefficients of Tokamak-plasmas are purely symmetric (i.e., $\tau_{\mu\nu}=\tau_{\nu\mu}$) \cite{balescu2}, in the sequel we shall set $f_{\mu\nu}=0$ so, $\tau_{\mu\nu}(X)=g_{\mu\nu}(X)$, and near equilibrium $\tau_{0\mu\nu}=L_{\mu\nu}$.

\subsection{Objective of the work}

\noindent In previous works, a macroscopic {\it Thermodynamic Field Theory} (TFT) for deriving the flux-force closure relations for systems out of Onsager's region has been proposed \cite{sonnino1}-\cite{sonnigiAIP}. The aim of the present work is to test the validity of this theory by studying some relevant examples of systems out of equilibrium. More specifically, we calculate the electron mass and energy losses in Tokamak-plasmas in turbulent regimes. We already noticed that Tokamak-plasmas are systems out of Onsager's region. Here we also highlight that the Onsager relations, which are at the basis of the so-called {\it Neoclassical theory}, strongly disagree with experimental data, and the disagreement amounts to several orders of magnitude. The discrepancy is even more pronounced in the case of magnetically confined plasmas in turbulent regimes. In this work, we use our model for computing the electron mass and heat losses in FTU (Frascati Tokamak Upgrade)-plasmas in turbulent transport regimes and compare the theoretical predictions with experimental data. We found that in this regime the theoretical predictions are in fairly good agreement with experiments. 

\subsection{Why study FTU-plasmas ?}

\noindent The Frascati Tokamak Upgrade (FTU) is a tokamak operating at Frascati, Italy. Building on the Frascati Tokamak experiment, FTU is a compact, high-magnetic-field tokamak ($B_{toroidal}$ = 8 Tesla). It began operation in 1990 and has since achieved operating goals of 1.6 MA at 8T and an average electron density greater than 4×1020 per cubic meter. The poloidal section of FTU is circular, with a limiter. Table~{\ref A} reports the main parameters for FTU-plasma (a detailed overview of FTU-plasma can be found in \cite{pucella}).
\begin{table}[t]
\centering
\footnotesize
\begin{tabular}{lr}
 \toprule
  \multicolumn{2}{c}{FTU}  \\
   \toprule
Major radius & 0.935\ meter	  \\
Minor radius & 0.35\ meter\\
Magnetic field & $<$\ 8\ Tesla\\
Plasma current & $<\ $1.6\ MA\\
Dicharge duration & 1.7\ sec\\
 \toprule
\end{tabular}
\caption{\footnotesize\textit{Main parameters for FTU.}}
\label{A}
\end{table}
\noindent {\it Low-confinement mode} (L-mode) and {\it High-confinement mode} (H-mode) are operating modes possible in toroidal magnetic confinement fusion devices - mostly Tokamaks. H-mode is when magnetically contained plasma is heated until it goes from a state of L-mode to H-mode state and becomes more stable and better confined. However, to allow plasma to reach the H-mode state, in addition to the Ohmic heating, auxiliary heating sources are necessary, such as radio-frequency heating, neutral-beam injection, etc. Therefore, H-mode Tokamak-plasmas are complex systems where exact analytical calculations are difficult to perform. FTU operates in L-mode and plasma is heated only by Ohmic heating. The main objective of this work is to compare the theoretical predictions provided by our model with the experimental data. FTU-plasma is the ideal fusion devise to perform analytical calculations as 
\begin{itemize}
\item{{\it FTU operates in L-mode};}
\item{{\it FTU-plasma is heated only by Ohmic heating};}
\item{{\it turbulence is relatively weak (due to the intense magnetic field)};}
\item{{\it dynamics is governed by only two thermodynamic forces, linked to the temperature and pressure gradients} (see the next section).}
\end{itemize}
These are the main reasons why we decided to study FTU-plasma.

\noindent The work is organized as follows. The basic aspects of our model are summarised in Section~\ref{theory}. Here, in order to set up vocabulary, we quickly introduce the definition of the {\it Space of the thermodynamic forces} and we describe the {\it Thermodynamic Covariance Principle} (TCP). Successively, in Section~\ref{transport} we investigate the simplest case of transport processes in Tokamak-Plasmas subject to two thermodynamic forces, linked to the thermal and pressure gradients. Subsection~\ref{Onsager} shows the (big) discrepancy between the predictions of the Neoclassical theory - which is based on the Onsager reciprocal relations - and experimental data (see Fig.~\ref{fig_test_Onsager}). In Subsection~\ref{TFT} we establish the mathematical framework to compute electronic mass and heat losses in turbulent Tokamak-plasmas subject to two thermodynamic forces. The agreement of our theoretical predictions with experimental data for FTU-plasmas can be found in Section~\ref{turbulence} (see Fig.~\ref{fig_Turb_comp}, in particular). We already anticipate that for computing losses in turbulent FTU-plasma we assumed that in this regime the transport coefficients are isotropic in the space of the thermodynamic forces. This is the first time that such an assumption, which is somewhat reminiscent of the Kolmogorov hypothesis for turbulent fluids (see, for instance, \cite{pope}, \cite{tennekes}), has been proposed in the literature. In Section~\ref{microscopic} we conciliate the macroscopic results with the microscopic ones. More specifically, the following results are obtained: 
\begin{itemize}
\item{by using the Prigogine's fluctuations theory for systems out of equilibrium \cite{prigogine2}, we derive the probability distribution function for the fluctuations of thermodynamic quantities;}
\item{we recover the expressions of the transport coefficients for turbulent plasmas obtained by phenomenological models (see \cite{bohm} and, for instance, \cite{gyrobohm1} and \cite{gyrobohm2}). In particular, we show that field theory, combined with the theorems established by the thermodynamics of irreversible processes, is able to evaluate the Bohm and gyro-Bohm coefficients. To the best of our knowledge, this is the first time that the expressions for these coefficients have been determined analytically.}
\end{itemize}
\noindent A brief discussion on the possibility to apply our results to the next generation of Tokamaks, as ITER or DTT, is reported in Section~\ref{perp}. Concluding remarks can be found in Section~\ref{conclusions},

\section{Modelling Transport Processes for Systems out of Onsager's Region}\label{theory}
\begin{figure}[h]
\centering\includegraphics[scale=.15]{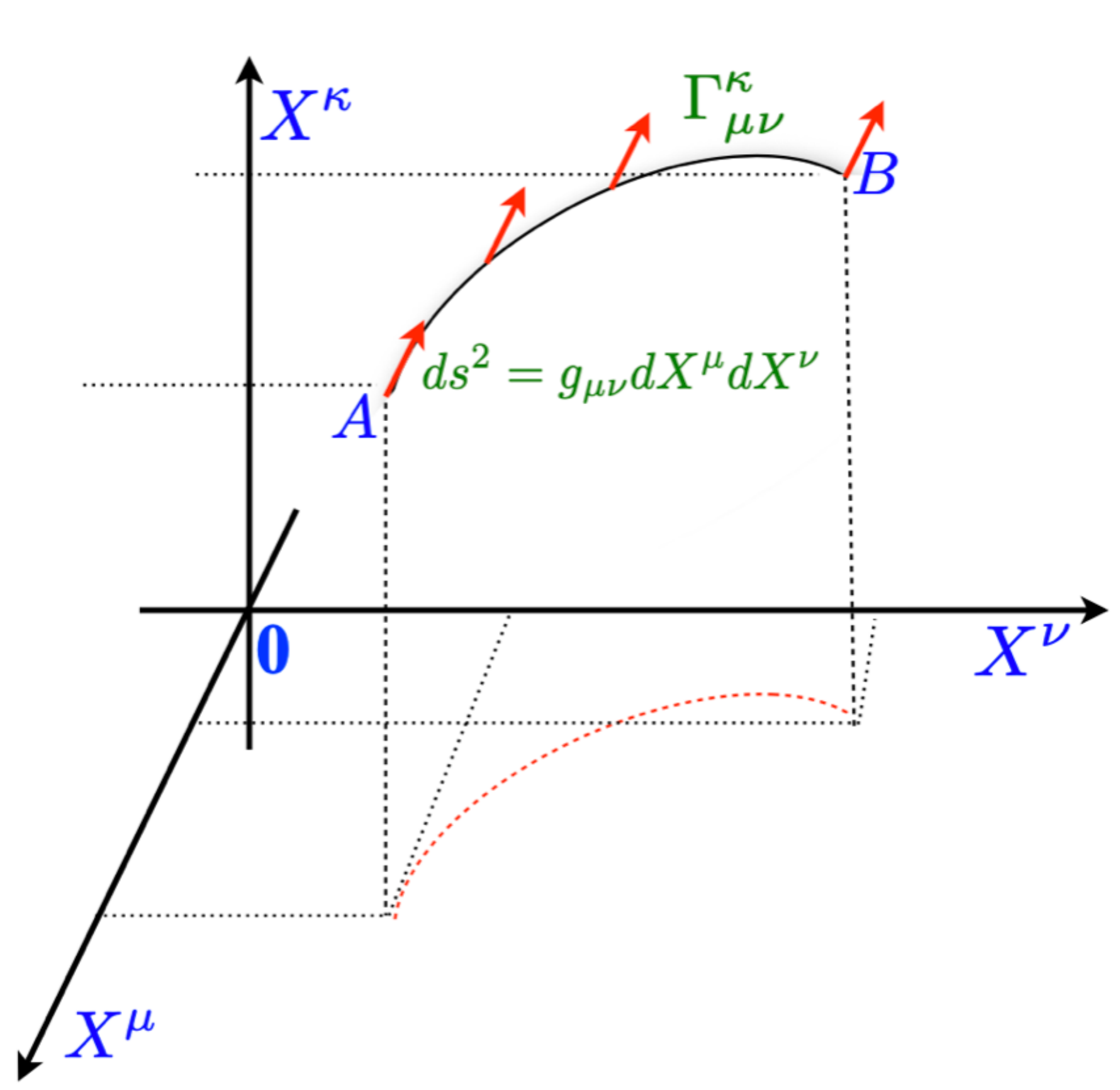}
\caption{{\bf The Space of Thermodynamic Forces}. The space is spanned by thermodynamic forces. The metric tensor is identified with the symmetric piece of the transport coefficients and the expression of the affine connection is provided by the {\it General Criterion of Evolution}. Note that for the second law of thermodynamics $ds^2=g_{\mu\nu}dX^\mu dX^\nu\geq 0$.}
\label{fig_thermodynamic_space}
\end{figure}

In our model, the equations to be satisfied by the transport coefficients are obtained by the thermodynamic theorems for systems out of equilibrium with the help of the tools provided by differential geometry. In order to establish the vocabulary and notations that shall be used in the sequel, we briefly recall the main definitions and concepts of this theory. First of all, we have to define the space where to carry out our calculations. We introduce the {\it Space of the thermodynamic forces} (or, simply, {\it the thermodynamic space}) defined as follows (see also Fig.~\ref{fig_thermodynamic_space}):
\begin{itemize}
\item{The space is spanned by the thermodynamic forces;}
\item{The metric tensor is identified with the symmetric piece $g_{\mu\nu}$ of the transport coefficients;}
\item{The expression of the affine connection $\Gamma_{mu\nu}^\lambda$ is determined by imposing the validity of the Glansdorff-Prigogine {\it General Criterion of Evolution} \footnote{The General Criterion of Evolution establishes that a thermodynamical system evolves towards a non-equilibrium steady-state in such a way that the expression $P=J_\mu\delta X^{\mu}$ is always a negative quantity, and $P=0$ at the steady-state \cite{prigogine3}, \cite{prigogine4}. }.}
\end{itemize}
\noindent

%%%%%%%%%%%%%%%%%%%%%%%%%%%%%%%%%%%%%%%%%%
%%%%%%%%%%%%%%%%%%%%%%%%%%%%%%%%%%%%%%%%%%
\noindent The model is based on three hypotheses: two constraints 1. and 2., and one assumption 3. We briefly recall these hypotheses.

\begin{enumerate}
\item {\textit{The thermodynamic laws and the theorems demonstrated for systems far from equilibrium must be satisfied.}}
\item{\textit{The validity of the Thermodynamic Covariance Principle (TCP) must be ensured}. 

\noindent The TCP stems from the concept of equivalent systems from the thermodynamic point of view. Thermodynamic equivalence was originally introduced by Th. De Donder and I. Prigogine \cite{prigogine1}, \cite{prigogine2}, \cite{degroot}. However, the De Donder-Prigogine definition of thermodynamic equivalence, based only on the invariance of the entropy production, is not sufficient to guarantee the equivalence character between two sets of thermodynamic forces and conjugate thermodynamic fluxes. In addition, it is known that there exists a large class of flux-force transformations such that, even though they leave unaltered the expression of the entropy production, they may lead to certain paradoxes \cite{Verschaffelt}, \cite{Davies}. The equivalent character of two alternative descriptions of a thermodynamic system is ensured if, and only if, the two sets of thermodynamic forces are linked with each other by the so-called \textit{Thermodynamic Covariant Transformations} (TCTs). The TCTs are the most general thermodynamic force transformations which leave unaltered both the entropy production $\sigma$ and the Glansdorff-Prigogine dissipative quantity $P$ defined as $P=J_\mu\delta X^{\mu}$. The TCT leads in a natural way to postulate the validity of the so-called \textit{Thermodynamic Covariance Principle} (TCP): {\it The nonlinear closure equations, i.e., the flux-force relations must be covariant under the Thermodynamic Covariant Transformations} (TCT).}
\item{\textit{Close to the steady states, the nonlinear closure equations can be derived by the principle of least action.}}
\end{enumerate}
\noindent The model, based on $1.$, $2.$, and $3.$, is referred to as the {\it Thermodynamical Field Theory} (TFT). The three hypotheses $1.$, $2.$, and $3.$ allow determining the nonlinear partial differential equations (PDEs) for transport coefficients $\tau_{\mu\nu}$. The explicit expression of the PDEs for a system subjected to $N$ independent thermodynamic forces are derived in \cite{sonninochaos}. In the present work, we shall limit ourselves to investigating the two-dimensional case i.e., when the system is subjected to two independent thermodynamic forces. In two dimensions, the {\it Riemannian curvature tensor} $B_{\mu\nu\lambda\kappa}$ has only {\it one independent component}, since all nonzero components may be obtained from $B_{0101}$. Equivalently, the curvature tensor may be written in terms of the scalar $B$ (see, for instance, \cite{weinberg})
\begin{equation}\label{2dim1}
B_{\lambda\mu\kappa\nu}=\frac{1}{2}B\left(g_{\lambda\kappa}g_{\mu\nu}-g_{\lambda\nu}g_{\mu\kappa}\right)
\end{equation}
\noindent So, $B$ alone completely characterizes the local geometry. However, since the Riemann tensor is constructed from the metric tensor, we may already deduce that even $g_{\mu\nu}$ can have only one independent component (the other components must depend on this). From Eq.~(\ref{2dim1}) we find $B_{\mu\nu}\equiv B_{\lambda\mu\kappa\nu}g^{\lambda\kappa}$ and $B\equiv B_{\lambda\mu\kappa\nu}g^{\lambda\kappa}g^{\mu\nu}$. So, we get
\begin{equation}\label{2dim2}
B_{\mu\nu}-\frac{1}{2}Bg_{\mu\nu}\equiv 0\qquad \forall\ g_{\mu\nu}
\end{equation}
\noindent Hence, the field equations obtained for $N\geq 3$ (i.e., when the system is subjected to at least three independent thermodynamic forces) are meaningless in two dimensions (see also \cite{brown}, \cite{collas}). In analogy with the works for $1+1$ gravity \cite{jackiw1}, \cite{teitelboim}, also in our case, when the system is subjected only to two independent thermodynamic forces the equation to be satisfied by the transport coefficients $g_{\mu\nu}$ is not tensorial but a {\it scalar differential equation} (see the forthcoming Eq.~(\ref{TPT15}). For a more detailed explanation see \cite{sonninochaos}). To sum up, in contrast to the case $N\geq3 $, where we obtain a differential equation for each component of $g_{\mu\nu}$, the case $N=2$ is degenerate (as the full Riemann tensor has only one non-trivial component), which leads to a unique differential equation that must be satisfied by the transport coefficients. It is useful now to recall the main result of the field theory based on a two-dimensional differential geometry: \textit{In a two-dimensional field theory, the metric tensor is conformal to a flat metric}. In practice, this means that for a two-dimensional case (but only in 2-dimensions) the transport coefficients, for a system out of Onsager's region, can always be brought into the form
\begin{equation}\label{TPT10}
g_{\mu\nu}(X)=\eta_{\mu\nu}\Lambda (X)
\end{equation}
\noindent with $\eta_{\mu\nu}$ denoting a (constant) flat metric. In literature, when two metrics are linked each with other by Eq.~(\ref{TPT10}) we say that $g_{\mu\nu}$ is {\it conformal to a flat metric} and $\Lambda(X)$ is referred to as the {\it conformal factor}. More in general, a conformal manifold is a manifold equipped with an equivalence class of metric tensors, in which two metrics $g_{\mu\nu}$ and ${\tilde g}_{\mu\nu}$ are equivalent if and only if
\begin{equation}\label{TPT11}
g_{\mu\nu}=\Lambda(X) {\tilde g}_{\mu\nu}
\end{equation}
\noindent where $\Lambda(X)$ is a real-valued smooth function defined on the manifold. Thus, a conformal metric may be regarded as a metric that is only defined {\it up to scale}. A conformal metric is conformally flat if the metric may be cast in the form (\ref{TPT10}). We have already mentioned that field theory, based on differential geometry, establishes that {\it a two-dimensional metric is always conformal to a flat metric}. Now, by requiring that the transport coefficients must tend to the Onsager matrix when the system is near equilibrium, we get
\begin{equation}\label{TPT13}
\eta_{\mu\nu}=L_{\mu\nu}
\end{equation}
\noindent So, 
\begin{equation}\label{TPT14}
g_{\mu\nu}=L_{\mu\nu}\exp\phi (X)
\end{equation}
\noindent where we have set $\Lambda(X)=\exp\phi (X)$ as $\Lambda(X)> 0$ $\forall X$ due to the second law of thermodynamics. Here, $\phi$ denotes a scalar field depending on the thermodynamic forces. In \cite{sonninochaos} we have determined the nonlinear partial differential equation that must be satisfied by the conformal field $\phi (X)$: 
\begin{equation}\label{TPT15}
L^{\mu\nu}\frac{\partial^2\phi}{\partial X^\mu\partial X^\nu}-\frac{4}{9\sigma_{L}}X^\mu X^\nu\frac{\partial^2\phi}{\partial X^\mu\partial X^\nu}-\frac{4}{9\sigma_{L}}X^\mu\frac{\partial\phi}{\partial X^\mu}+\frac{5}{9\sigma_{L}}\left(X^\mu\frac{\partial\phi}{\partial X^\mu}\right)^2=0
\end{equation}
\noindent with $\sigma_L\equiv L_{\mu\nu}X^\mu L^\nu$. Eq.~(\ref{TPT15}) has to be solved with the appropriate boundary conditions. Notice that the equilibrium condition requires $\phi(0)=0$. Eq.~(\ref{TPT15}) is the equation that determines the only surviving Riemannian component $B_ {0101}$. In Thermodynamical Field Theory, Eq.~(\ref{TPT15}) is analogous to Liouville’s equation in Riemannian (or pseudo-Riemannian) geometry \cite{jackiw1}, \cite{jackiw2} and \cite{cavaglia}). It is also worth recalling that, as pointed out in \cite{becchi}, in 2-dim, even if {\it locally} the metric is conformal to a flat metric, {\it globally} the ensemble of charts - commonly called {\it atlas} - may not be conformal to a flat metric (e.g., consider the case of a sphere where the charts overlap at the poles generating a non-trivial metric at those points) \cite{becchi}. The forthcoming Subsection~\ref{TFT} is devoted to the study of transport in Tokamak-plasmas subject to Temperature and pressure gradients. This task will be accomplished by solving PDE~(\ref{TPT15}) for plasma in a turbulent regime (FTU-plasma in collisional regime is treated in \cite{sonninochaos}). We conclude this Section by mentioning that the proposed model has been applied successfully to unimolecular triangular chemical reactions (i.e., three isomerisations take place), to materials subjected to temperature and electric potential gradients, and to nonlinear Hall-effect (see \cite{sonnino2} and \cite{sonnigi8}-\cite{sonnigi10}).

\section{Transport Processes in Tokamak-Plasmas subject to Temperature and Pressure Gradients}\label{transport} 
\subsection{Tokamak-plasma in Onsager's region: the Neoclassical Theory}
Let us consider a Tokamak-plasma subject to two thermodynamic forces $X^1$and $X^2$. The expression for the entropy production $\sigma$ reads
\begin{equation}\label{TPT1}
\sigma=J_1 X^1+J_2 X^2
\end{equation}
\noindent with $J_1$ and $J_2$ denoting the thermodynamic fluxes conjugate to $X^1$ and $X^2$, respectively. When the system is in the linear region of thermodynamics (Onsager's region), the transport matrix reads
\begin{equation}\label{TPT2a}
L_{\mu\nu}=\begin{pmatrix}
{\tilde\sigma}_\parallel & {\tilde\alpha}_\parallel\\
{\tilde\alpha}_\parallel & {\tilde\kappa}_\parallel
\end{pmatrix}
\end{equation}
\noindent with ${\tilde\sigma}_\parallel$, ${\tilde\alpha}_\parallel$, and ${\tilde\kappa}_\parallel$ denoting the {\it dimensionless (parallel) electronic conductivity}, {\it dimensionless (parallel) thermoelectric coefficient}, and the {\it dimensionless (parallel) thermal conductivity}, respectively \cite{balescu2}. Particle and energy electronic losses are given, respectively, by the expressions
\begin{eqnarray}\label{TPT2b}
&&\left<{\bf J}^{e(1)}\right>_r=-K_e\left<\frac{\beta_0}{B}\left(1-\frac{B^2}{\beta_0^2}\right)X^1\right>\\
&&\left<{\bf J}^{e(3)}\right>_r=K_e\left<\frac{\beta_0}{B}\left(1-\frac{B^2}{\beta_0^2}\right)X^2\right>\nonumber
\end{eqnarray}
\noindent with
\begin{eqnarray}\label{TPT3}
&&\beta_0\equiv \left<B^2\right>^{1/2}\\
&&K_e=\frac{B_\xi}{B_\theta}\left(\frac{1}{\Omega_{e0}}\tau_e\right)\nonumber
\end{eqnarray} 
\noindent Here $<\cdots>$ denotes the operation {\it average magnetic surface}, and $B_\xi$ and $B_\theta$ are the magnetic components in the $\xi$ direction (the toroidal angle) and in the $\theta$ direction (the poloidal angle), respectively. $\Omega_e$ and $\tau_e$ denote the electronic Larmor frequency and the electronic collision time, respectively. In case of Tokamak-plasmas, the thermodynamic forces coincide with the {\it dimensionless (parallel) generalized forces}. Notice that $X^1$ and $X^2$ are not measurable quantities. The expressions for the conjugate thermodynamic fluxes read \cite{balescu2}
\begin{equation}\label{TPT4}
J_\mu=\left(\frac{B}{\beta_0}-\frac{\beta_0}{B}\right)K_e\begin{pmatrix}
g_\rho^{(1)P} \\
-g_\rho^{e(3)}
\end{pmatrix}
\end{equation}
\noindent with
\begin{eqnarray}\label{TPT5}
&&g_\rho^{(1)P}=-\tau_e\left(\frac{T_e}{m_e}\right)^{1/2}\left(1+\frac{P_i}{P_e}\right)\frac{\nabla_\rho P}{P}\\
&&g_\rho^{e(3)}=-\sqrt{\frac{5}{2}}\tau_e\left(\frac{T_e}{m_e}\right)^{1/2}\frac{\nabla_\rho T_e}{T_e}\nonumber
\end{eqnarray} 
\noindent Here, $\rho$ is the radial spatial component, and $T_e$ and $m_e$ denote the temperature and mass of the electron, respectively. $P_i$, $P_e$ are the ionic and electronic pressure respectively and $P=P_i+P_e$. Notice that the thermodynamic fluxes are measurable quantities. The Onsager entropy production $\sigma_{Onsager}$ is defined as
\begin{equation}\label{TPT6}
\sigma_{Onsager}=L^{\mu\nu} J_\mu J_\nu
\end{equation}
\noindent The {\it strategy} adopted by the neoclassical theory is very simple: 
\begin{itemize}
\item{The expressions for the thermodynamic forces are obtained by inverting the Onsager relations: $X^\mu=L^{\mu\nu}J_\nu$ (with $\mu,\nu=1,2$);}
\item{The expressions for the electron mass and energy losses are successively obtained by using Eq.~(\ref{TPT2b}).}
\end{itemize}

\subsection{Comparison between the Neoclassical Predictions and Experimental Data}\label{Onsager}
\begin{figure}[h]\centering
\centering\centering\includegraphics[scale=.15]{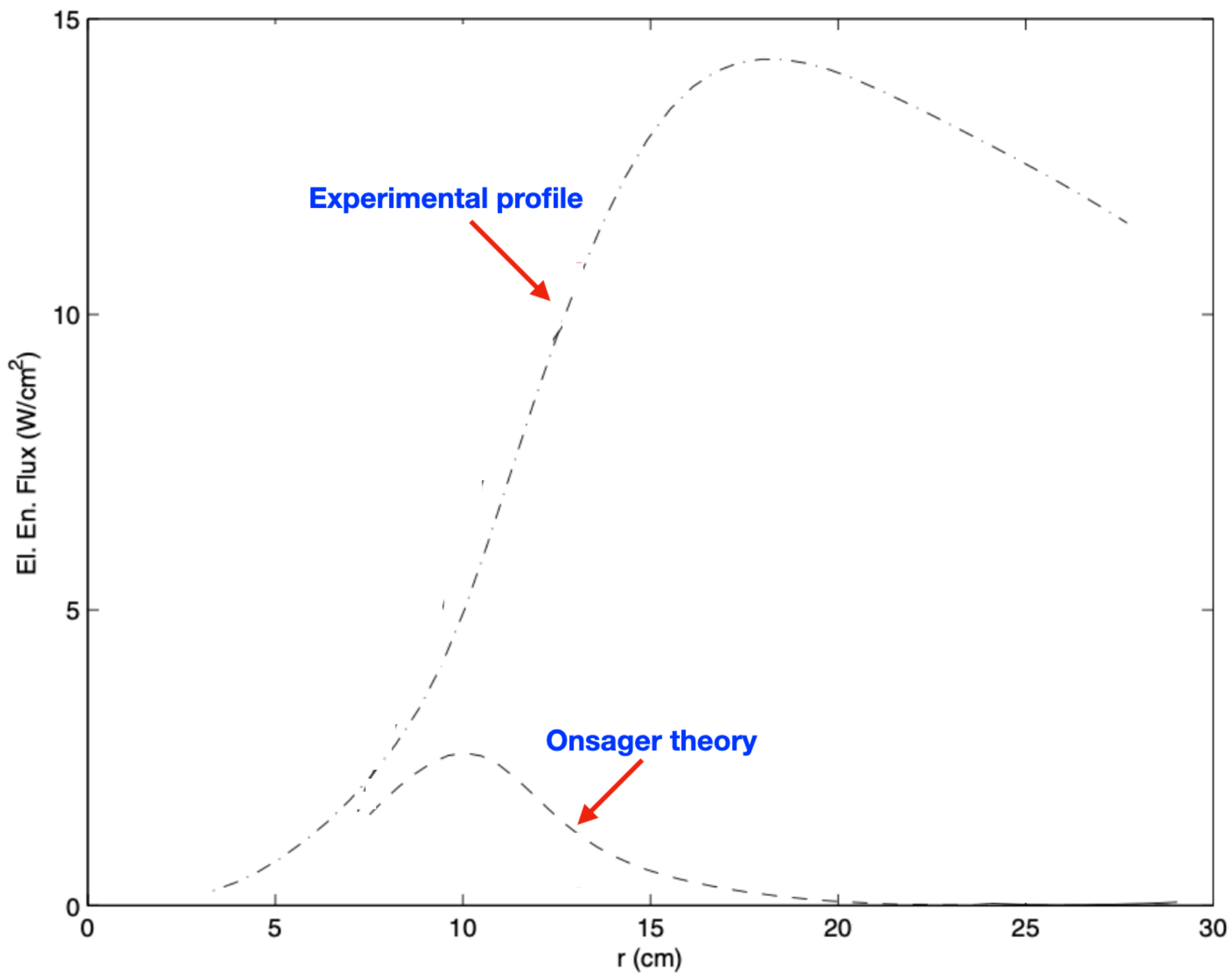}
\caption{{\bf Energy electron loss in FTU-plasma: Comparison between the Neoclassical prediction (Onsager's theory) and experimental data}. The discrepancy between theory and experimental data is of an order of magnitude.}
\label{fig_test_Onsager}
\end{figure}

Table~{\ref B} compares the main parameters related to the three Tokamaks currently working FTU, JET (Joint European Torus), and under construction ITER (International Thermonuclear Experimental Reactor). Here, $B_T=$ Toroidal magnetic field at the major radius, $I=$ Plasma current, $R=$Major radius of the Tokamak, and $P_{aux}$= Auxiliary power.
\begin{table}[t]
\centering
\footnotesize
\begin{tabular}{lrrrr}
 \toprule
  & $B_T$ & I(MA)& R(m)&${\rm P}_{{\rm aux}}$(MW)  \\
   \toprule
FTU	&	8	&	1.6	&0.93 & 4.6 \\
JET	&	3.5	&	6	& 3.0 & 20\\
ITER	&	5.7	&	21	& 8.1 & 300\\
 \toprule
\end{tabular}
\caption{\footnotesize\textit{Comparison between the parameters of FTU, JET and ITER.}}
\label{B}
\end{table}
\noindent As we can see, the intensity of the magnetic field is noticeably higher in FTU than in the other two reactors. So, in FTU-plasmas, turbulence is more contained as a high value of the magnetic field tends to {\it freeze} the turbulence of charged particles (ions and electrons). In addition, the value of the external auxiliary power is much less in FTU than in the other two reactors. Finally, FTU-plasma is subjected only to two thermodynamic forces and this simplifies considerably the expression of the entropy production. So, FTU is the ideal experimental device where calculations can be performed quite easily. This is the reason why we decided to test the validity of our model by investigating FTU-plasmas. In general, in Tokamak-plasmas the electron energy flux maybe even be 10,000 times greater that the theoretical predictions obtained by the Neoclassical theory, which is constructed by adopting the linear (Onsager) relations. Fig.~\ref{fig_test_Onsager} shows the comparison between the Onsager theory predictions and experimental data for energy electron loss in FTU-plasmas.  

%%%%%%%%%%%%%%%%%%%%%%%%%%%%%%%%%%%%%%%%%%
%%%%%%%%%%%%%%%%%%%%%%%%%%%%%%%%%%%%%%%%%%
\noindent The discrepancy, both quantitatively and qualitatively, between these two profiles is evident: the large difference between the two energy electron loss profiles denotes that FTU-plasma is out of the Onsager region. Furthermore, Fig.~\ref{fig_test_Onsager} shows that close to the edge of the Tokamak the concavity of Onsager's profile is directed upwards while the experimental profile shows a concavity facing downwards. 

\subsection{Tokamak-plasma out of Onsager's region}\label{TFT}

When the plasma is out of the Onsager region, the flux-force relation may be brought into the form
\begin{equation}\label{TPT7}
J_\mu=g_{\mu\nu} X^\nu
\end{equation}
\noindent The {\it strategy} for computing the electron mass and energy losses remains the same: 
\begin{itemize}
\item{By Eq.~(\ref{TPT4}) we get the expression for the thermodynamic forces by inverting Eq.~(\ref{TPT7}): $X^\mu=g^{\mu\nu}J_\nu$ with $J_\nu$ given by Eq.~(\ref{TPT4}) and $\mu=1,2$;}
\item{The expressions for the electronic mass and energy losses are successively obtained by using Eq.~(\ref{TPT2b}).}
\end{itemize}
\noindent However, we do not know the expressions of the transport coefficients when the system is out of Onsager's region. For this, we have to set up a model able to determine these expressions. Several theories, mainly based on phenomenological approaches, have been proposed in the literature; we cite among them the most popular ones: the Bohm and the gyro-Bohm phenomenological transport models (see, for example, \cite{bohm} and \cite{gyrobohm1}-\cite{gyrobohm2}, respectively). The Bohm model predicts that the rate of diffusion is linear with temperature and inversely linear to the strength of the confining magnetic field strength $B$:
\begin{equation}\label{TPT8}
\chi_{Bohm}=\frac{1}{16}\frac{k_B T}{e_\iota B}=C_{Bohm}\frac{k_B T}{e_\iota B}
\end{equation}
\noindent where $e_\iota$ is the elementary charge of species $\iota$ ($\iota=$electron, ion) and $k_B$ is the Boltzmann constant, respectively. The gyro-Bohm diffusion model $\chi_{gyroBohm}$ reads (see, for example, \cite{gyrobohm1})
\begin{equation}\label{TPT9}
\chi_{gyroBohm}=C_{gyroBohm}\frac{k_B T}{e_\iota \beta_0}\frac{\rho_{\iota 0}}{a}
\end{equation}
\noindent Here, $C_{gyroBohm}$ is a constant factor, which has to be adjusted so that the simulation results well reproduce experimental observations. $\rho_{\iota 0}$ and $a$ are the ion Larmor radius of species $\iota$ evaluated at the averaged magnetic field $\beta_0$ and the minor radius of the Tokamak, respectively. Hence, in Eqs~(\ref{TPT8}) and (\ref{TPT9}), the Bohm and gyro-Bohm coefficients are determined experimentally or by numerical simulations. In the forthcoming Section~\ref{microscopic} we shall see that these coefficients can be evaluated by field theory combined with the thermodynamics of irreversible processes (TIP).

%%%%%%%%%%%%%%%%%%%%%%%%%%%%%%%%%%%%%%%%
\section{Testing the validity of the PDE~(\ref{TPT15}) for FTU-plasma in Turbulent Regime}\label{turbulence}
The main purpose of the series of works we produce is to test the validity of the PDE~(\ref{TPT15}). To this aim, we compare the theoretical predictions with experimental data provided by the EUROfusion Consortium in Frascati (Rome-Italy) for FTU-plasmas. Two distinct regimes need to be studied: 
\begin{itemize}
\item{FTU-plasma in collisional transport regime;}
\item{FTU-plasma in turbulent regime.}
\end{itemize}
\noindent Experiments show that the first case corresponds to the plasma located in the core of the Tokamak. The latter case corresponds to the plasma near the edge of the Tokamak.

\subsection{FTU-plasma in fully collisional regime: plasma in the core of the Tokamak}
\begin{figure}[h]
\centering\includegraphics[scale=.65]{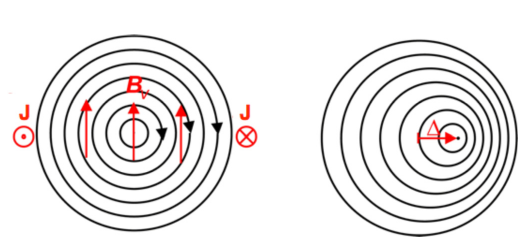}
\caption{{\bf The Shafranov shift}. In tokamak-plasmas, the plasma pressure leads to an outward shift $\Delta$ of the center of the magnetic flux surfaces. ${\bf J}$ indicates the direction of the electric current that flows inside the plasma. Note that the poloidal magnetic field increases and the magnetic pressure can, then, balance the outward force \cite{sonninochaos}.}
\label{fig_Shafranov_shift}
\end{figure}

\begin{figure}[h]
\centering\includegraphics[scale=.50]{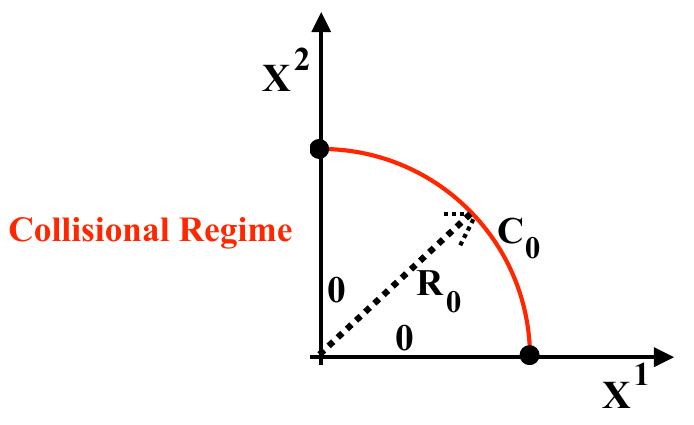}
\caption{{\bf Boundary conditions for $\phi$ for collisional FTU-plasmas}. The validity of Onsager's relations is ensured by imposing that $\phi$ vanishes along the axes. In addition, we have to impose that there are no privileged directions for very large values of the thermodynamic forces. This condition is satisfied by imposing, in the first quadrant, $\phi(r=R_0,\theta)=c_0\neq0$.}
\label{fig_BC1}
\end{figure}

This regime has already been investigated in \cite{sonninochaos}. For easy reference, and to prevent the reader from getting lost, we report here, very briefly, the main results obtained, referring the reader to \cite{sonninochaos} for more details. In \cite{sonninochaos} experiments have been performed in a zone of the Tokamak where the turbulent effects are almost {\it frozen}. In our calculations, we have also taken into account the Shafranov-shift (which is not negligible in FTU-plasmas). The physical explanation of the Shafranov-shift is briefly sketched in Fig.~(\ref{fig_Shafranov_shift}). 
%%%%%%%%%%%%%%%%%%%%%%%%%%%%%%%%%%%%%%%%%%
%%%%%%%%%%%%%%%%%%%%%%%%%%%%%%%%%%%%%%%%%%
\noindent To compare the theoretical predictions with experimental data we have specified the boundary conditions for Eq.~(\ref{TPT15}). It is worth mentioning how these have been obtained:
\begin{enumerate}[{\bf a.}]
\item First of all, we have to satisfy the Onsager condition. Hence, the solution should vanish at the origin of the axes, i.e., $\phi(\mathbf{0}) =0$;

\item Experimental evidences show that, in a pure collisional regime, the pure effects (such as Fourier's law, Fick's law, etc.) are very robust laws. So, we have to impose $\phi(X^1=0,X^2)=\phi(X^1,X^2=0)=0$;

\item There are no privileged directions when the thermodynamic forces tend to infinity (or for very large values of the thermodynamic forces). In other words, $\phi(r=R_0,\theta)=const.\equiv c_0\neq 0$ on the arc of a circle of radius $R_0$ (with $R_0$ very large). Here, $(r,\theta)$ denotes the polar coordinates: $r=({X^1}^2+{X^2}^2)^{1/2}$ and $\theta=\arctan(X^{2}/X^{1})$. The boundary conditions, in the case of FTU-plasmas in a fully collisional regime are depicted in Fig.~(\ref{fig_BC1}).
\end{enumerate}                                   
%%%%%%%%%%%%%%%%%%%%%%%%%%%%%%%%%%%%%%%%%%
%%%%%%%%%%%%%%%%%%%%%%%%%%%%%%%%%%%%%%%%%%
\noindent Parameters $R_0$ defines the minimal circle enclosing the solution area. Its value has been determined by following the procedure shown in \cite{sonninochaos}. The Dirichlet boundary condition $c_0$ is determined such that the thermodynamic forces $X^1$ and $X^2$ maximize the electron heat loss. In other words, $c_0$ is the solution of the equation
\begin{equation}\label{TPT16a}
\frac{d}{dc}\left<{\bf J}^{e(3)}\right>_{(c, r=r_M(c))}=0
\end{equation}
\noindent with $r_M(c)$ solution of the equation
\begin{equation}\label{TPT16b}
\frac{d}{dr}\left<{\bf J}^{e(3)}\right>_{(c, r)}=0
\end{equation}
\noindent Eqs~(\ref{TPT16a}) and (\ref{TPT16b}) have been solved numerically; we get (approximatively) $c_0\simeq -4.5$. We have then solved the PDE~(\ref{TPT15}) in the first quadrant. After having obtained the solution in the first quadrant, by using the Schwartz principle \cite{CourantHilbert1}, we reconstructed the entire solution which is valid for the whole circle. We found that the non-linear theoretical profile provided by our model is much better in agreement with experimental data than Onsager's prediction. However, near the edge of the Tokamak, there is still a discrepancy between theory and experiments both quantitatively and qualitatively (as the concavity is upwards in the theoretical profile while it is downwards in the experimental curve - see the forthcoming Section and Fig.~\ref{fig_Turb_comp}).

\subsection{FTU-plasma in turbulent regime: plasma close to the edge of the Tokamak}\label{turbulent_edge}
\begin{figure}[h]
\centering\includegraphics[width=5.0cm]{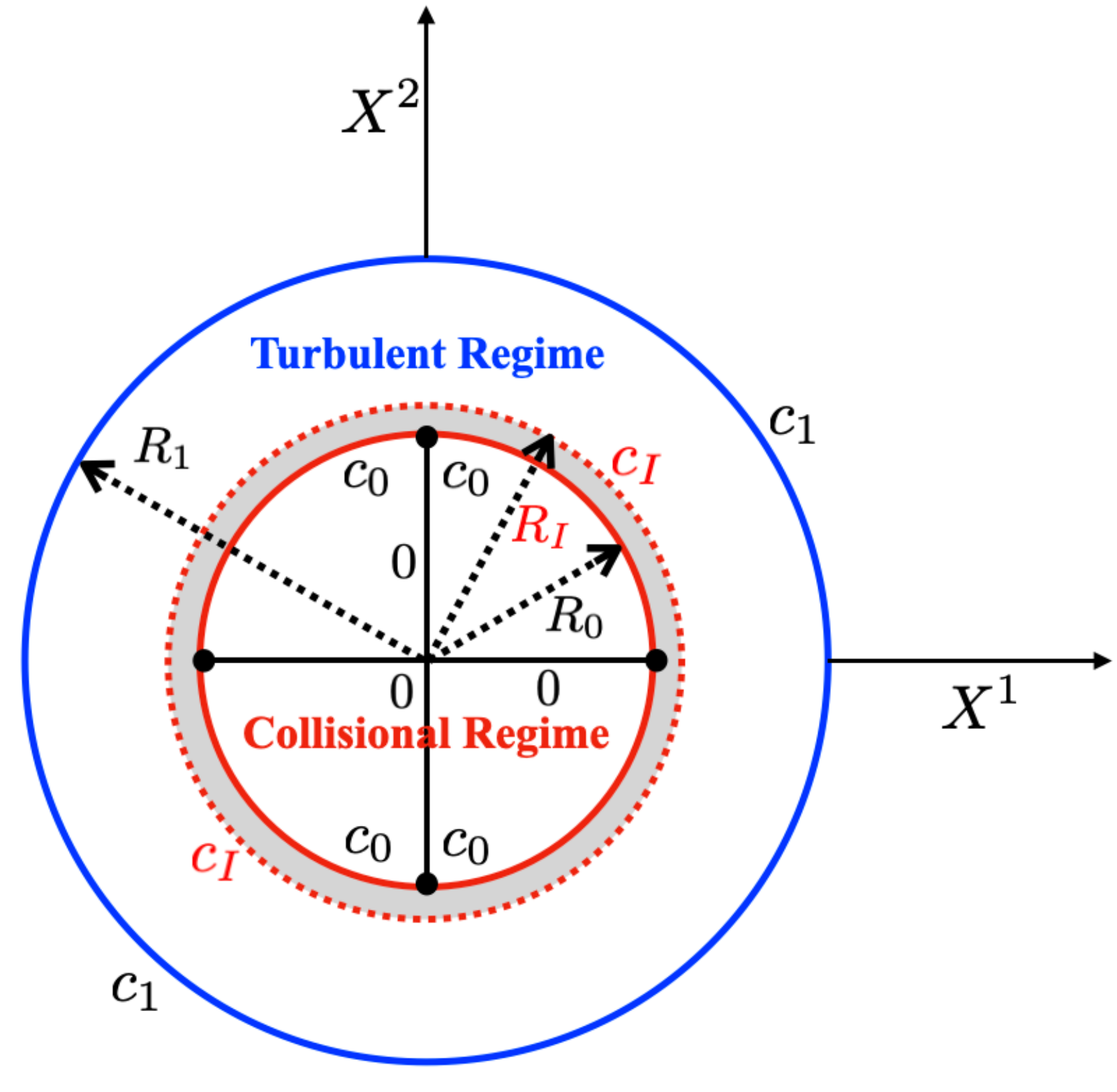}
\caption{{\bf Boundary conditions for $\phi$ for FTU-plasmas in collisional and in turbulent regimes}. In the turbulent zone, plasmas is far from equilibrium. So, in the annulus, the Onsager relations are no longer valid. Of course, we have always to impose that there are no privileged directions for very large values of the thermodynamic forces (say for $r=R_1$). This condition is satisfied by imposing, in the first quadrant, $\phi (r = R_1,\theta)=c_1\neq 0 $. Combining the solutions with these boundary conditions with those specified for the collisional FTU plasmas we obtain the graph of the solution of Eq.~(\ref{TPT15}) for both the collisional and the turbulent regimes. The grey ring finger corresponds to the {\it intermediate zone}.}
\label{fig_BC2}
\end{figure}

We already mentioned that discrepancies are observed near the edge of the Tokamak, where transport is mainly dominated by turbulence. Turbulent transport in Tokamak-plasmas is one of the main open issues that remain to be addressed in order to achieve good energy confinement in fusion devices. Generally, investigations are based on quasilinear gyro-kinetic modeling of turbulence. However, it has recently been shown that field theories, combined with theorems established by the thermodynamics of irreversible processes (TIP), are able to provide significant contributions to the solution of this problem. Furthermore, as we shall see, they allow evaluation of the Bohm and gyro-Bohm coefficients. Since Eq.~(\ref{TPT15}) has been derived without neglecting any term present in the balance equations (i.e., the energy, mass, and momentum balance equations), it is quite natural to propose this equation as a good candidate also for describing transport in two-dimensional turbulent systems. To analyze the electron heat loss for FTU-plasmas in the turbulent zone we have to specify the appropriate boundary conditions (BCs). In the turbulent zone, plasma is far from equilibrium and the Onsager relations are no longer valid. However, the problem is significantly simplified by assuming that very far from equilibrium there are no privileged directions in the space of thermodynamic forces. This condition is satisfied by imposing, in the first quadrant, $\phi(r=R_1,\theta) = c_1 \neq 0$. with $R_1/R_0\gg 1$. The boundary conditions for FTU-plasmas in collisional and in turbulent regimes are depicted in Fig.~(\ref{fig_BC2}). Notice that between the two regions, the collisional region and the turbulent annular, there is an intermediate zone depicted in grey.
%%%%%%%%%%%%%%%%%%%%%%%%%%%%%%%%%%%%%%%%%%
%%%%%%%%%%%%%%%%%%%%%%%%%%%%%%%%%%%%%%%%%%
Establishing the correct boundary conditions for the intermediate zone and finding the solution of Eq.~(\ref{TPT15}) satisfying these BCs is a very difficult task. In the grey annular area, the dynamic of the plasma is governed by both collisions and turbulence. In this work we investigate FTU-plasmas in the turbulent regime and in \cite{sonninochaos} we analyzed the collisional situation. Hence, in our approach, these two regimes are studied separately by limiting ourselves to connecting the solutions between these two regimes by simply drawing a red dashed curve.

\subsection{Transport Processes in Tokamak-Plasmas in Isotropic Turbulent Regime}

Let us now solve Eq.~(\ref{TPT15}) in the edge of the tokamak. Even in two dimensions (in the space of thermodynamic forces) this task remains difficult to accomplish. In order to facilitate the study, we assume the validity of the following {\bf Isotropic hypothesis}: {\it The transport coefficients have the same intensity regardless of the direction in space of the thermodynamic forces} \footnote{This assumption is somewhat reminiscent of {\it Kolmogorov's hypothesis of local isotropy} \cite{kolmogorov} (see also, for instance, \cite{pope}, or \cite{tennekes}). However, in our case isotropy refers to the space of thermodynamic forces (and not to the ordinary space) and homogeneity assumption is not required.}.

\noindent Isotropy means uniformity in all orientations. We recall that $\sigma_L$ is defined as
\begin{equation}\label{T1}
\sigma_L=L_{\mu\nu}X^\mu X^\nu
\end{equation}
\noindent So, by performing an orthogonal coordinate transformation
\begin{equation}\label{T2}
{X'}^\lambda=A^\lambda_\kappa X^\kappa
\end{equation}
\noindent with $A^\lambda_\kappa$ satisfying the relation:
\begin{equation}\label{T4}
A^\alpha_\lambda L^{\lambda\kappa}A^\beta_\kappa={\rm I}^{\alpha\beta}
\end{equation}
\noindent with ${\rm I}^{\alpha\beta}$ denoting the components of the identity matrix\footnote{Note that linear transformations of the thermodynamic forces are allowed as this class of transformations belong to the group TCT \cite{sonnino1} and, since $L^{\mu\nu}$ is a symmetric positive definite matrix, it is always possible to determine $A^\lambda_\kappa$ such that Eq.~(\ref{T4}) is satisfied.}, $\sigma_L$ may be cast into the form 
\begin{equation}\label{T3}
\sigma_L={X^{'1}}^2+{X^{'2}}^2=R^2
\end{equation}
\noindent Hence, assuming the isotropic hypothesis is equivalent to postulating that the transport coefficients depend only on $\sigma_L$ \footnote{$\sigma_L$ has the dimension of entropy production, but it differs from it. The entropy production of the system is $\sigma=g_{\mu\nu}X^\mu X^\nu$.}. Finally, the isotropic hypothesis can be equivalently expressed as: {\it When plasmas is in turbulent regime, the transport coefficients depend only on the bilinear expression of the entropy $\sigma_L$}.

\noindent In the sequel, we shall apply the assumption that in the edge of the FTU, plasmas are in {\it isotropic turbulent regime}. Note that Eq.~(\ref{TPT14}) allows to link of the entropy production $\sigma$ of plasma with the bilinear form $\sigma_L$. We get
\begin{equation}\label{T5}
\sigma=\sigma_L\Lambda(X)
\end{equation}
\noindent Summing up, in our problem we have three expressions having the dimension of the entropy production: $\sigma$, $\sigma_L$, and $\sigma_{Onsager}$. It is easily checked that, {\it in two dimensions}, these expressions are linked with each other by the relation
\begin{equation}\label{T5}
\sigma=\left(\sigma_L\sigma_{Onsager}\right)^{1/2}
\end{equation}
\noindent with
\begin{equation}\label{T6}
\sigma=g_{\mu\nu}X^\mu X^\nu\quad ;\quad \sigma_L=L_{\mu\nu}X^\mu X^\nu\quad ;\quad \sigma_{Onsager}=L^{\mu\nu}J_\mu J_\nu\nonumber
\end{equation}
\noindent {\bf Remark}

\noindent We may adopt two hypotheses, {\it a priori} equally acceptable:
\begin{enumerate}[{\bf i)}]
\item $ \Lambda(X)=\Lambda(\sigma_L)$
\item $\Lambda(X)=\Lambda (\sigma)$
\end{enumerate}

\noindent Assumption {\bf i)} states that the {\it transport coefficients are isotropic in the space of the thermodynamic forces} while assumption {\bf ii)} states that the {\it transport coefficients depend only on the entropy production}. We have analyzed both cases. In this work, we report only the results relating to the first case {\bf i)}. The procedure for treating case {\bf ii)} is quite similar to case {\bf i)}, albeit it is a little more complicated.
\subsection{Solution of Eq.~(\ref{TPT15})}
\begin{figure}[h]
\centering\includegraphics[width=10.0cm]{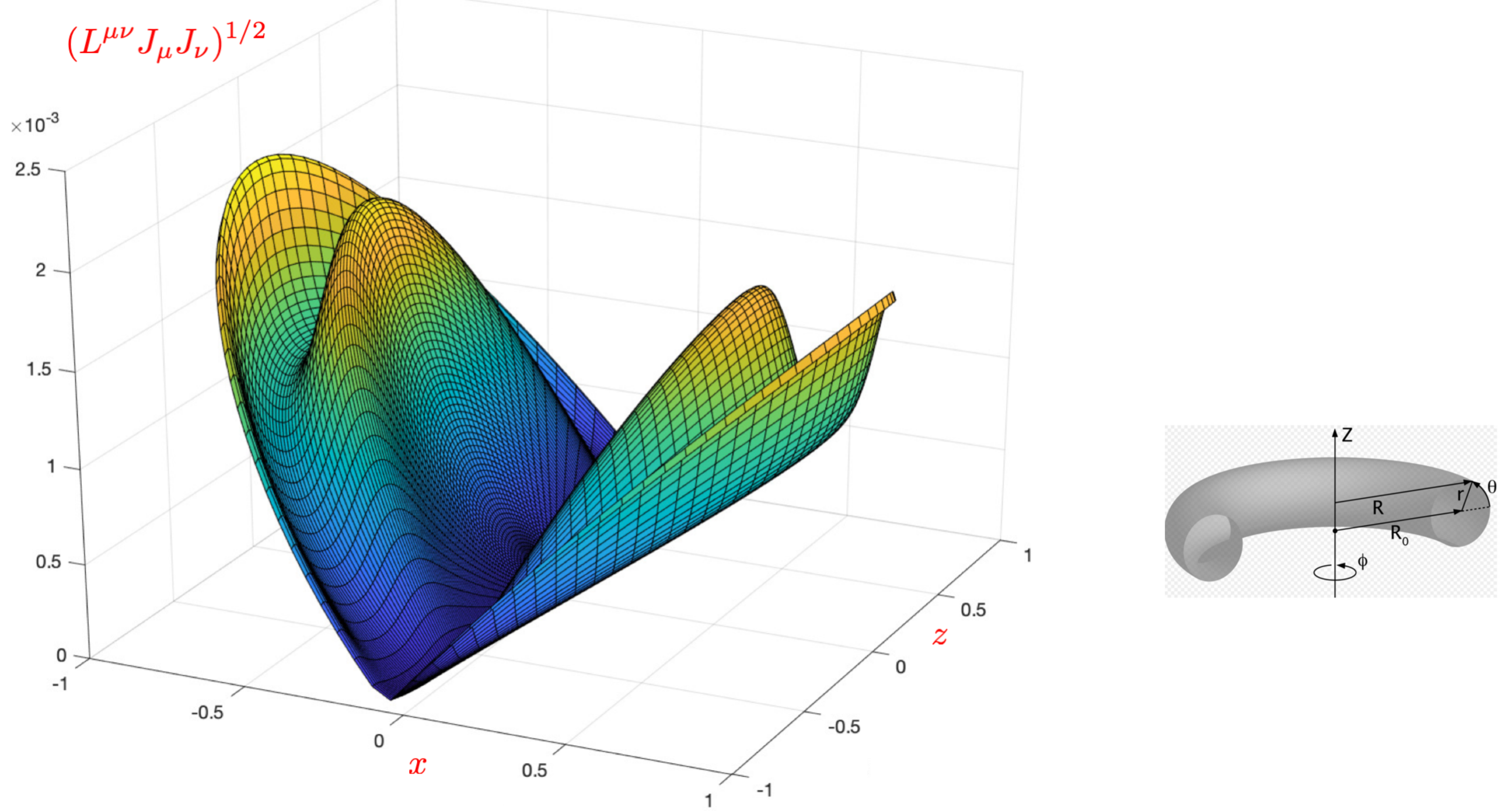}
\caption{{\bf Behaviour of (the square root of) Onsager's entropy production}. The square root of the Onsager entropy production is the r.h.s term of Eq.~(\ref{T7}). Coordinates $x$ and $z$ are linked to the coordinates $(r,\theta)$ by the relations $x= r/a\cos\theta$ and $z= r/a\sin\theta$.}
\label{fig_known_coeff}
\end{figure}

\begin{figure}[h]
\centering\includegraphics[width=10.0cm]{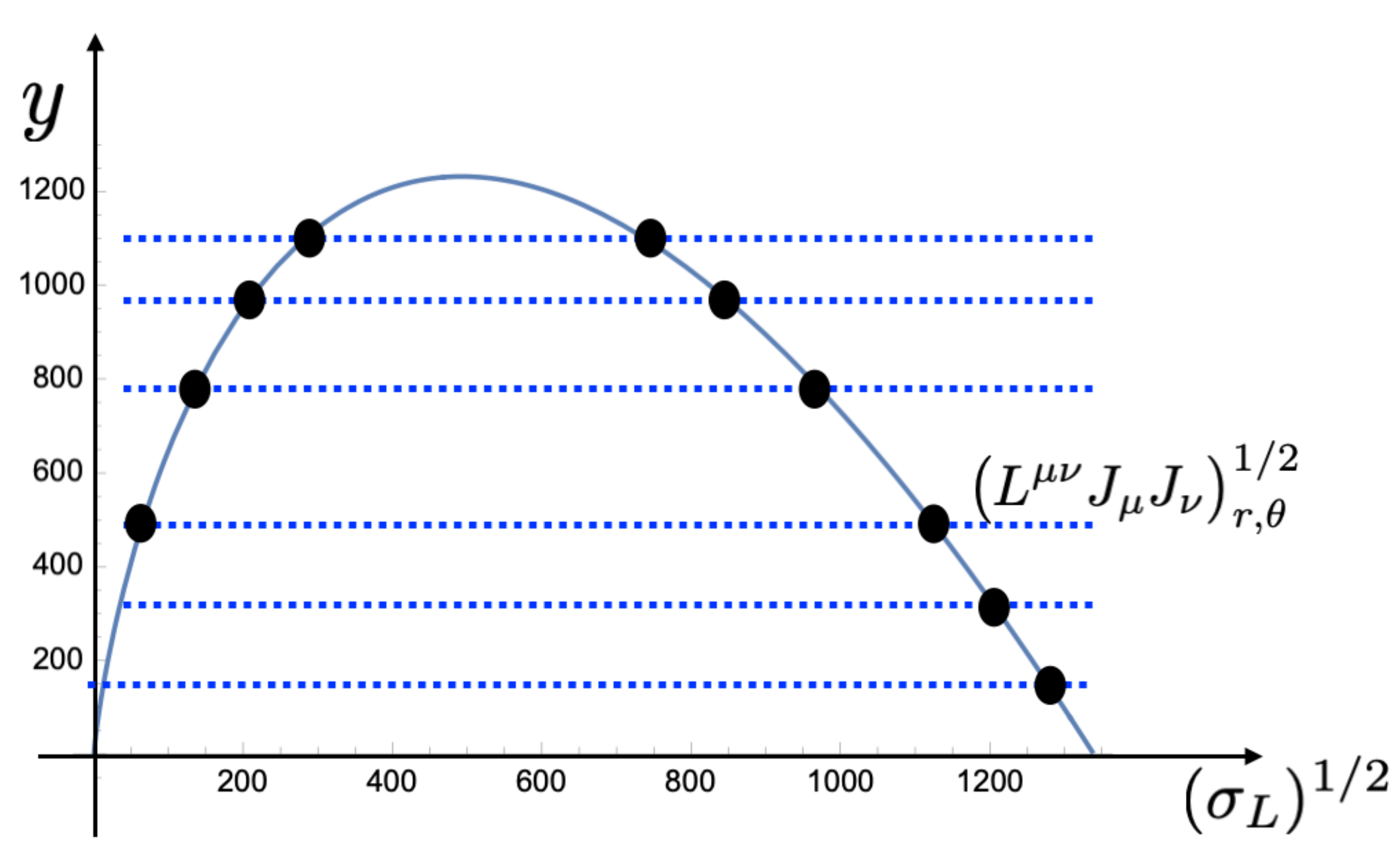}
\caption{{\bf determination of the solution of Eq.~(\ref{T7})}. This solution is the intersection between the curve $y=2\alpha\sigma_L^{1/2}\log(\sigma_L^{1/2})+\beta\sigma_L^{1/2}$ and the straight line $y=(L^{\mu\nu}J_\mu J_\nu)_{r,\theta}^{1/2}$ by varying the values of $(r,\theta)$ in the range $0\leq r\leq a$ and $0 \leq\theta \leq 2 \pi$.}
\label{fig_sol_Th_Turb}
\end{figure}

We can check that, under the isotropic assumption (i.e., $\Lambda(X)=\Lambda(\sigma_L)$), the solution of Eq.~(\ref{TPT15}), subjected to the boundary conditions $\Lambda(\sigma_L)\mid_ {R=R_I}=\Lambda({\sigma_{{}_{LI}}})$ and $\Lambda'({\sigma_{L}})\mid_{R=R_I}=\Lambda'({\sigma_{{}_{LI}}})$, reads:
\begin{equation}\label{T6}
\Lambda(\sigma_L)=\alpha\log(\sigma_L)+\beta
\end{equation}
\noindent where prime denotes the derivative with respect to $\sigma_L$, and $\alpha$ and $\beta$ are two constants that are determined by the boundary conditions. From Eq.~(\ref{T6}) we obtain the {\it limit values} of $R_1$ and $c_1$: $R_1=\exp(-\beta/\alpha)$ and $c_1=-\infty$\footnote{These values are easily obtained by setting $\Lambda(\sigma_L)=\exp(\phi)=0$.}. Recalling that $\Lambda=\sigma/\sigma_L$, we finally get
\begin{equation}\label{T7}
2\alpha\sigma_L^{1/2}\log(\sigma_L^{1/2})+\beta\sigma_L^{1/2}=(L^{\mu\nu}J_\mu J_\nu)^{1/2}
\end{equation}
\noindent where Eq.~(\ref{T5}) has been taken into account. Our aim is to compute the electron mass and energy losses. To carry out this task we must, first of all, determine the thermodynamic forces (see Eqs~(\ref{TPT2b})). The forces are derived by solving Eq.~(\ref{T7}) and by the relation
\begin{equation}\label{T8}
X^\mu=\left(\frac{\sigma_L}{L^{\kappa\lambda}J_\kappa J_\lambda}\right)^{1/2}L^{\mu\nu} J_\nu
\end{equation}
\noindent 
\begin{itemize}
\item{{\bf Determination of the constants $\alpha$ and $\beta$}}
\end{itemize}
\noindent We immediately see that the values of $\alpha$ and $\beta$ are linked to the values of $\Lambda$ and $\Lambda'$ at the boundary of the intermediate zone (see Fig.~\ref{fig_BC2}) by the relations:
\begin{eqnarray}\label{T9}
&&\alpha=\sigma_{\!\!{}_{LI}}\Lambda'(\sigma_{\!\!{}_{LI}})\\
&&\beta=\Lambda(\sigma_{\!\!{}_{LI}})-\sigma_{\!\!{}_{LI}}\Lambda'(\sigma_{\!\!{}_{LI}})\log (\sigma_{\!\!{}_{LI}})\quad{\rm with} \quad \sigma_{\!\!{}_{LI}}=\Lambda^{-1} (\sigma_{\!\!{}_{LI}})\sigma_{\!\!{}_{I}}\nonumber
\end{eqnarray}
\noindent Hence, the determination of the values of the constants $\alpha$ and $\beta$ requires the knowledge of the values of the constants $c_I$ and $R_I$. For this, we have to solve Eq.~(\ref{TPT15}) in the intermediate region. We did not accomplish this (very complex) task. We have assumed that the intermediate region is sufficiently narrow to consider that these values are very close to those taken by the collisional solution at the boundary. So, at $R_I \simeq R_0$, we have $ c_I\simeq c_0=-4.5$ (and $\Lambda(\sigma_L)_{R=R_I}\simeq \exp{(-4.5)}$), $\sigma _ {R=R_I}\simeq\sigma_{R=R_0}=1.5\ 10^{-5}$ (see Fig.~\ref{fig_entropy_coll}), and $\Lambda'(\sigma_{\!\!{}_{LI}})\simeq \Lambda'(\sigma_{\!\!{}_{L}})_{R=R_0}=-925.75$. We finally get
\begin{equation}\label{T10}
\alpha=-1.25\quad {\rm and}\quad \beta=18
\end{equation}
\noindent Fig.~\ref{fig_known_coeff} shows the behaviour of the term $(L^{\mu\nu}J_\mu J_\nu)_{x,z}^{1/2}$ vs the coordinates $x=r/a\cos\theta$ and $z=r/a\sin\theta$, with $a$ denoting the minor radius of the Tokamak. 
%%%%%%%%%%%%%%%%%%%%%%%%%%%%%%%%%%%%%%%%%%
%%%%%%%%%%%%%%%%%%%%%%%%%%%%%%%%%%%%%%%%%%
\noindent Fig.~(\ref{fig_sol_Th_Turb}) illustrates the strategy adopted for solving Eq.~(\ref{T7}): we have determined the points of intersection (the black ones in Fig.~(\ref{fig_sol_Th_Turb})) between the curve $y=2\alpha\sigma_L^{1/2}\log (\sigma_L^{1/2})+\beta\sigma_L^{1/2}$ and the straight line $y=(L^{\mu\nu}J_\mu J_\nu)_{r,\theta}^{1/2}$ for a given value of $(r,\theta)$. The expression for $\sigma_L(r,\theta)$, function of $r$ and $\theta$, is obtained by varying the values of $(r, \theta)$ in the range $0\leq r\leq a$ and $0 \leq\theta \leq 2 \pi$.
%%%%%%%%%%%%%%%%%%%%%%%%%%%%%%%%%%%%%%%%%%
%%%%%%%%%%%%%%%%%%%%%%%%%%%%%%%%%%%%%%%%%%

\begin{figure}[h]
\centering\includegraphics[width=5.0cm]{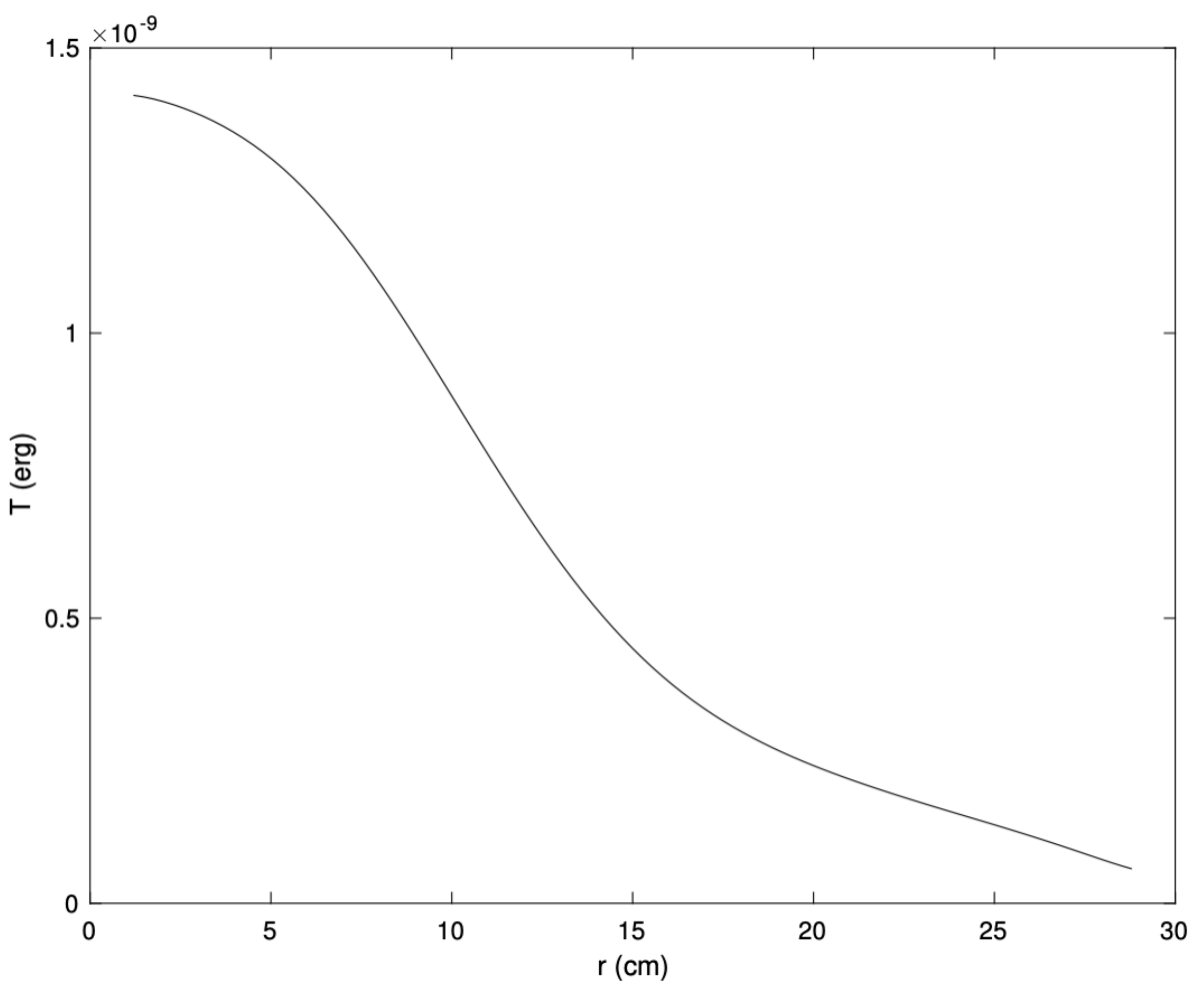}
\caption{{\bf Experimental Temperature profile of FTU-plasma at the moment of the shot}. These data have been provided by Massimo Marinucci  (ENEA Fusion Division in Frascati - Rome).}
\label{fig_Temperature}
\end{figure}

\begin{figure}[h]
\centering\includegraphics[width=5.0cm]{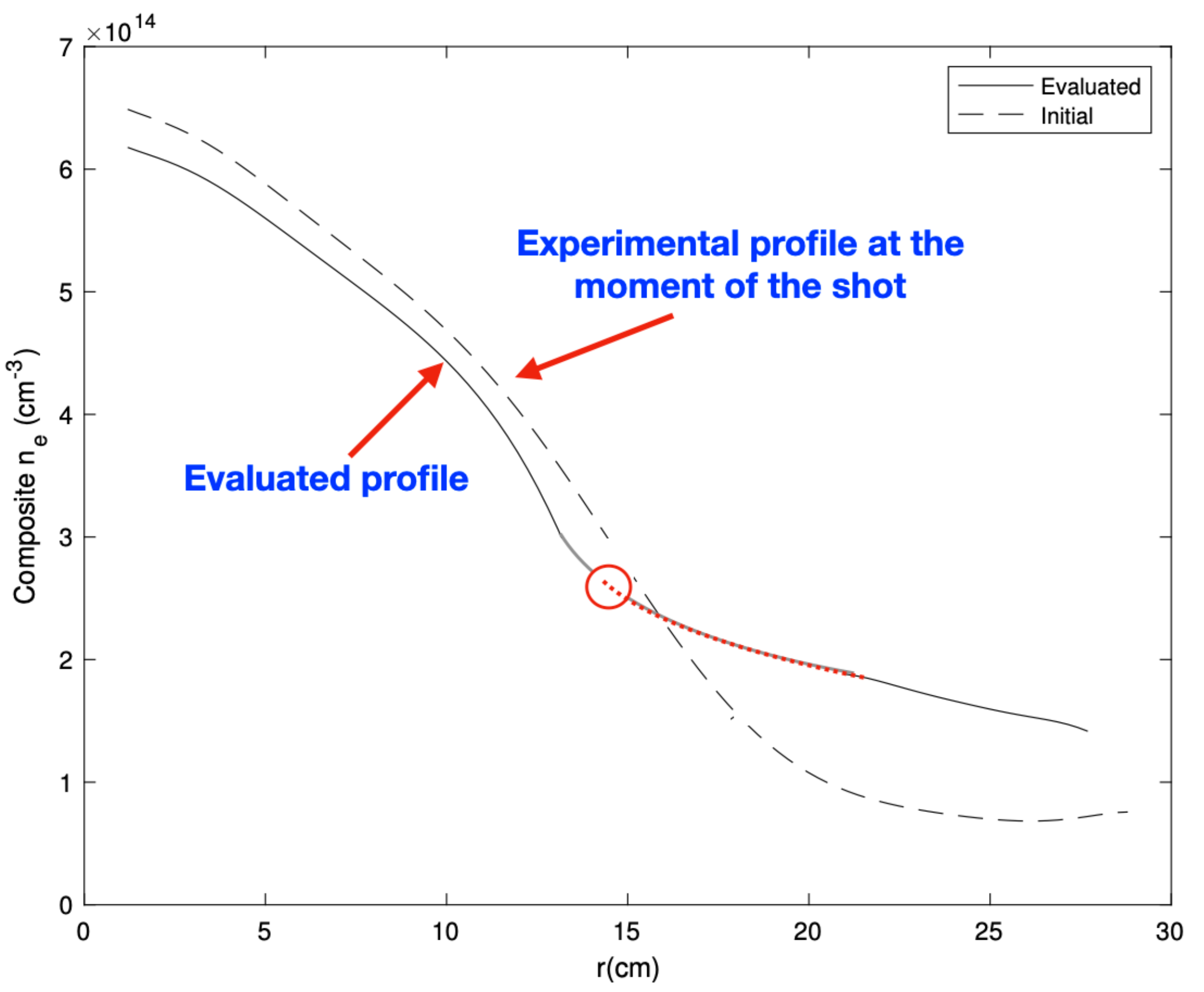}
\caption{{\bf The experimental profile (dashed line) refers to the moment of the shot}. The full line is the stationary density profile which has been obtained by imposing the electron number conservation and the continuity of the composite function (the collisional profile with the turbulent one) with its derivative. The experimental data at the moment of the shot have been provided by Massimo Marinucci  (ENEA Fusion Division in Frascati - Rome).}
\label{fig_Density}
\end{figure}

\begin{figure}[h]
\centering\includegraphics[width=10.0cm]{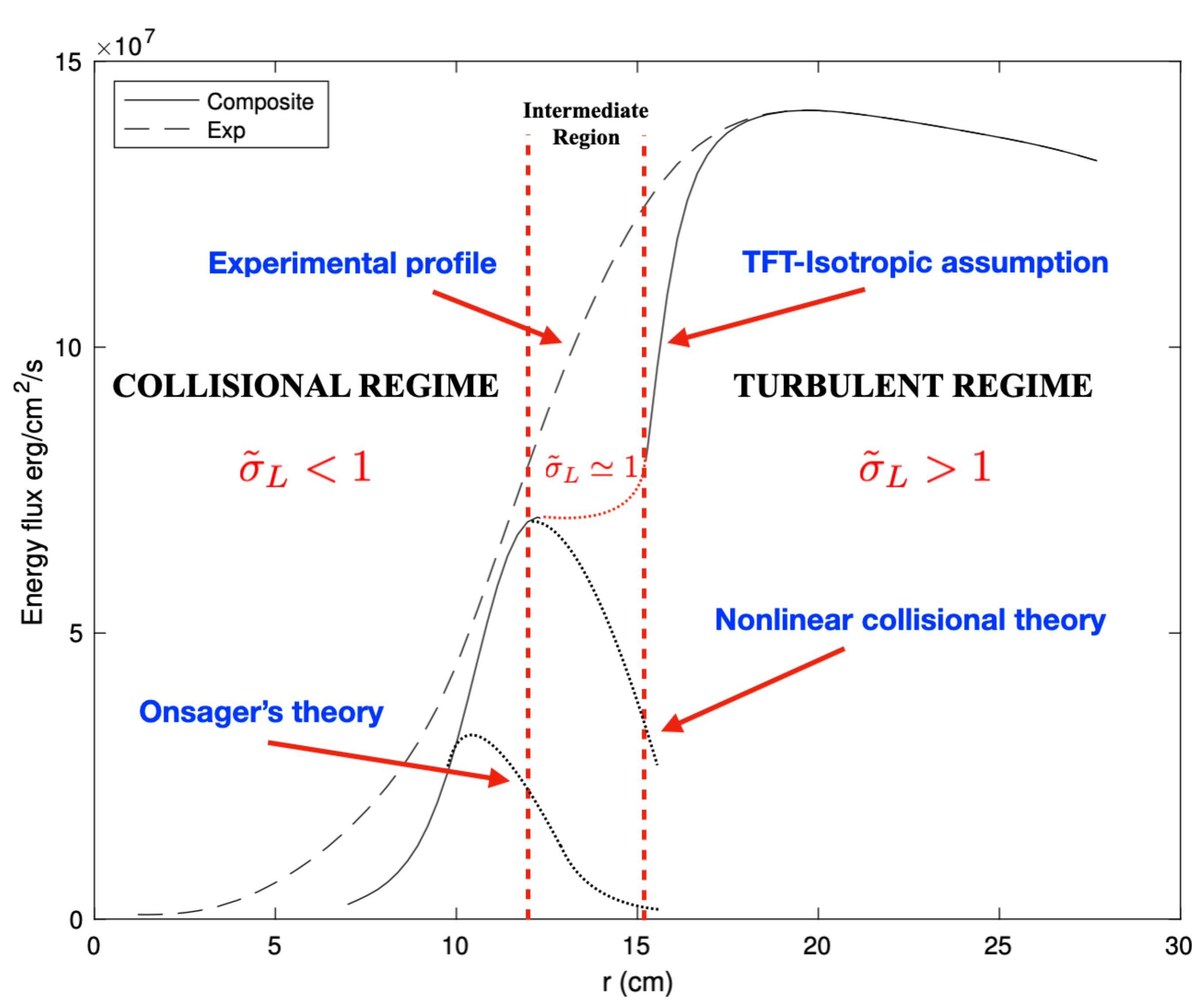}
\caption{{\bf Comparison between the theoretical predictions and the experimental data}. $\tilde\sigma_L\equiv 10^{3}\sigma_L$. The dashed line corresponds to the experimental data. The black full line is the composition of the two solutions the collisional one and the turbulent one. The dashed red line corresponds to the intermediate region. The dot black profiles are the Neoclassical (Onsager) prediction and the nonlinear collisional prediction. The experimental data have been provided by Massimo Marinucci  (ENEA Fusion Division in Frascati - Rome).}
\label{fig_Turb_comp}
\end{figure}

\noindent We conclude this Subsection by mentioning that in Tokamak-plasmas it is possible to keep the temperature profile stable even after the shot. However, it is impossible to avoid the diffusion of the species (ions and electrons) towards the edges of the Tokamak, and this is due to various types of {\it drifts} and the {\it toroidal configuration} of the Tokamak (see, for instance, \cite{balescu2}). Fig.~(\ref{fig_Temperature}) and Fig.~(\ref{fig_Density}) show the experimental Temperature and density profiles. Fig.~(\ref{fig_Density}) illustrates the density profile at the moment of the shot (dashed line) and the evaluated stationary density profile right after the shot (full line). This latter profile has been obtained by imposing two conditions:
\begin{itemize}
\item{the electron number conservation};
\item{the continuity of the function, and of its derivative, of the composite profiles obtained by joining the solution in the collisional regime (plasma in the core of the Tokamak) with the turbulent one (plasma close to the edge of the Tokamak). The red dotted line corresponds to the intermediate region (see Fig. ~ \ref{fig_BC2}).}
\end{itemize}
%%%%%%%%%%%%%%%%%%%%%%%%%%%%%%%%%%%%%%%%%%
%%%%%%%%%%%%%%%%%%%%%%%%%%%%%%%%%%%%%%%%%%
%%%%%%%%%%%%%%%%%%%%%%%%%%%%%%%%%%%%%%%%%%
%%%%%%%%%%%%%%%%%%%%%%%%%%%%%%%%%%%%%%%%%%
\noindent Fig.~\ref{fig_Turb_comp} shows the comparison between the theoretical predictions and the experimental data for FTU-plasma. The dashed line is the experimental profile. The black full line are the solutions of Eq.~(\ref{TPT15}) for FTU-plasma in the collisional regime (the lower part of the curve) and in the turbulent regime (the top of the curve). The profile in the turbulent region has been obtained by assuming that the transport coefficients are isotropic in the space of the thermodynamic forces. The dashed red line corresponds to the intermediate region. This figure also shows the neoclassical (Onsager) prediction and the non-linear collisional profile obtained by our model (the black dotted lines).
%%%%%%%%%%%%%%%%%%%%%%%%%%%%%%%%%%%%%%%%%%
%%%%%%%%%%%%%%%%%%%%%%%%%%%%%%%%%%%%%%%%%%

\noindent $\bullet$ {\bf Physical origin of the turbulence}

\noindent The following question arises spontaneously: {\it "Can a fluid (or plasma) subject only to sufficiently high temperature and pressure gradients reach a turbulent regime? In this case, what physical mechanism brings the system into this regime}?" 

\noindent The answer to this question is \cite{fauve}: {\it "Yes. The mechanism by which a turbulent regime is reached when a fluid or a plasma is submitted to a temperature or pressure gradient is the hydrodynamic instability or a sequence of hydrodynamic instabilities. Usually, after each instability, more modes are excited and the flow becomes more complex. If the instability is subcritical, the flow can jump from a laminar to a turbulent state without the need for a sequence of instabilities. This occurs in parallel flows generated by a pressure gradient. Note however that the definition of turbulent flow is not a precise one. It is enough to have a flow which is chaotic in time and involves a wide range of exciting spatial scales"}.

\section{From the Macroscopic to the Microscopic Description - Some Hints}\label{microscopic}

\begin{figure}[h]
\centering\includegraphics[width=7.0cm]{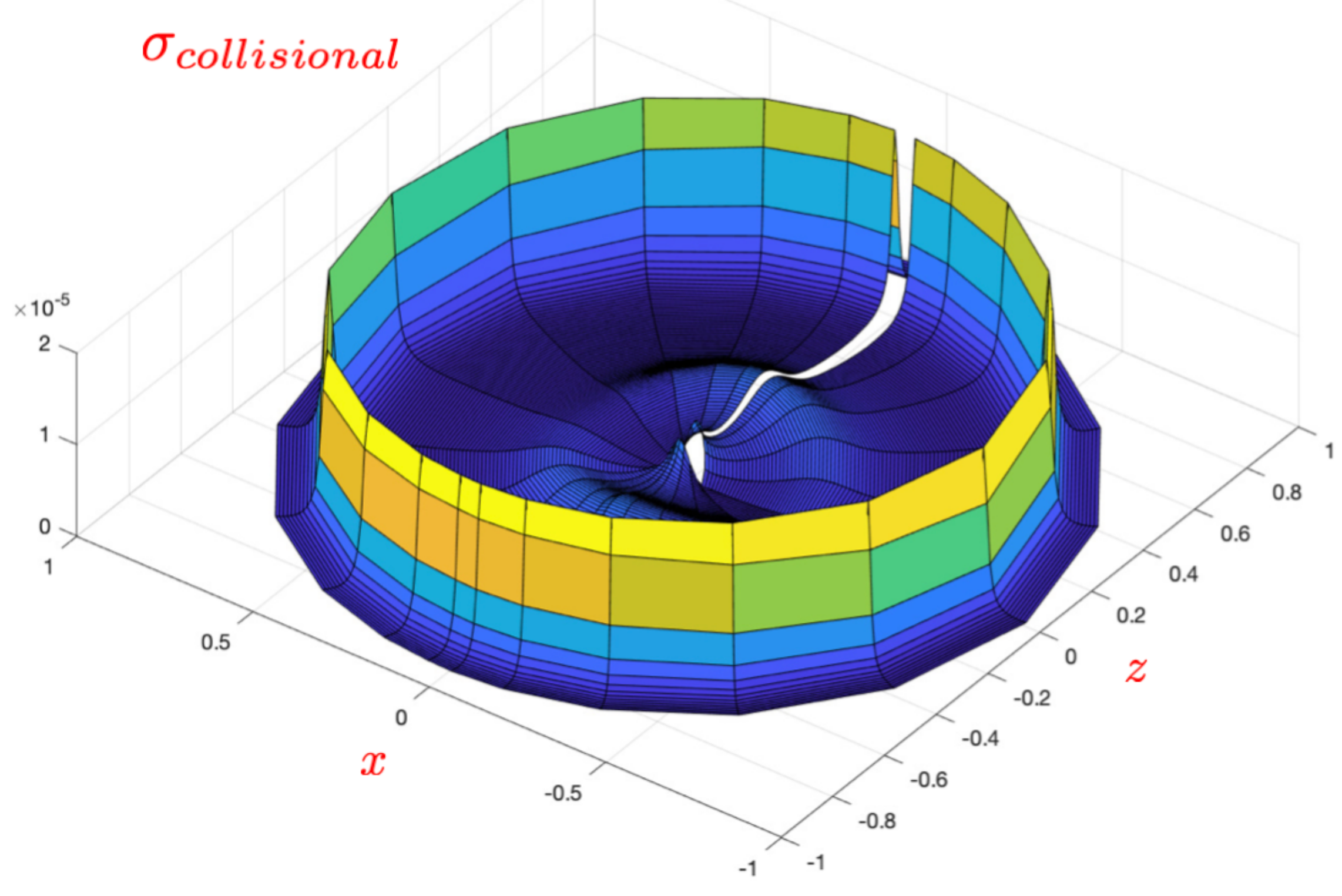}
\caption{{\bf Entropy production for FTU-plasma in the collisional regime}.}
\label{fig_entropy_coll}
\end{figure}

\begin{figure}[h]
\centering\includegraphics[width=7.0cm]{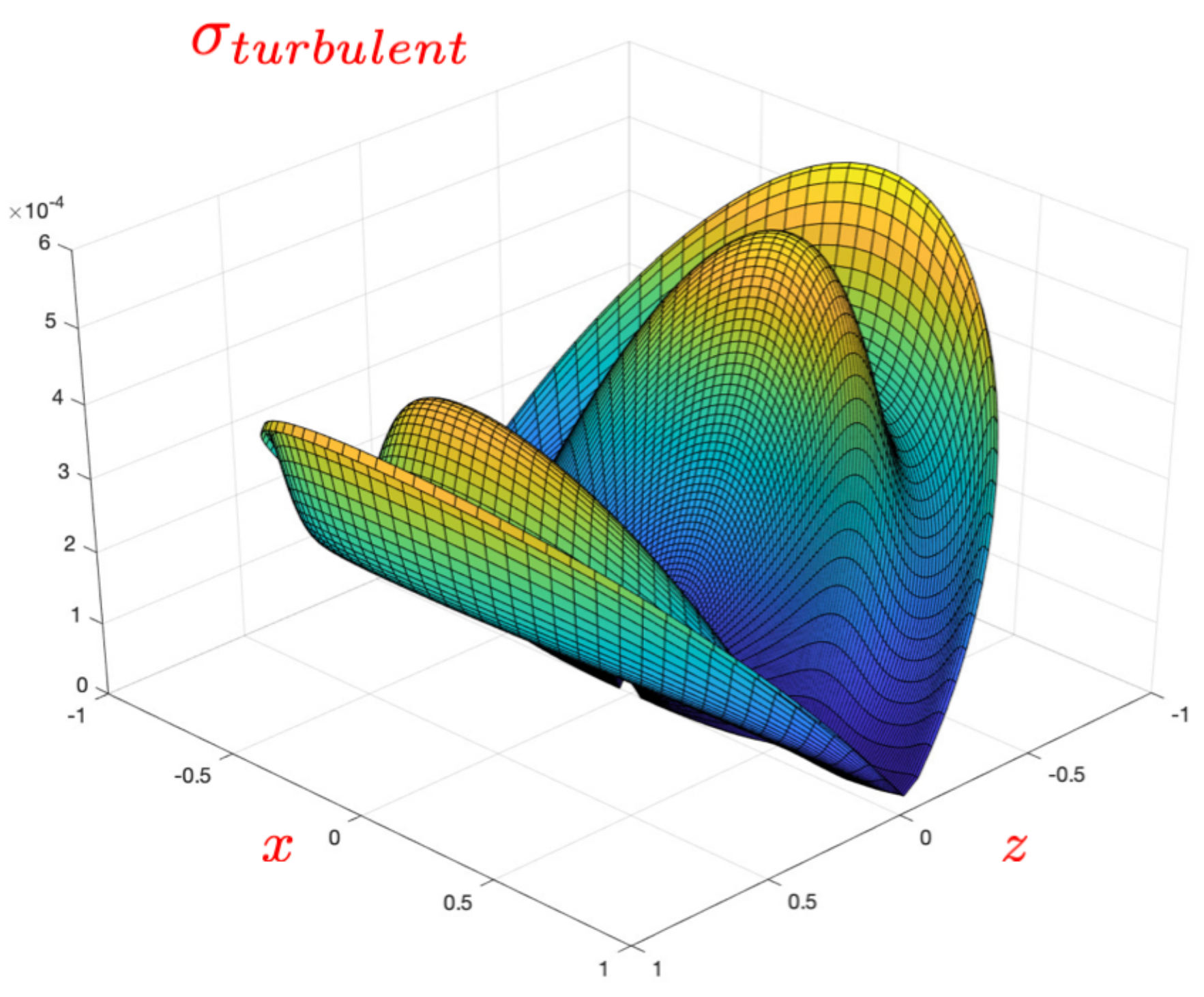}
\caption{{\bf Entropy production for FTU-plasma in turbulent regime}.}
\label{fig_entropy_turb}
\end{figure}

 A well-founded microscopic explanation of the validity of the linear phenomenological laws was developed by Onsager in 1931 \cite{onsager1}, \cite{onsager2}. Onsager's theory is based on three assumptions: i) {\it The probability distribution function for the fluctuations of thermodynamic quantities} (Temperature, pressure, degree of advancement of a chemical reaction etc.) {\it is a Maxwellian} ii) {\it Fluctuations decay according to a linear law} and iii) {\it The principle of the detailed balance} (or the microscopic reversibility) {\it is satisfied}.  Onsager showed the equivalence of Eqs (\ref{I2}) and (\ref{I3}) with the assumptions i)-iii) [assumption iii) allows deriving the {\it reciprocity relations} $\tau_{0\mu\nu}=\tau_{0\nu\mu}$]. The Onsager theory of fluctuations starts from the Einstein formula linking the probability of a fluctuation, $\mathcal W$, with the entropy change, $\Delta S$, associated with the fluctuations from the state of equilibrium
\begin{equation}\label{Mm1}
\mathcal{W}=W_0\exp[\Delta S/k_B]
\end{equation}
\noindent In Eq.~(\ref{Mm1}), $k_B$ is the Boltzmann constant and $W_0$ is a normalization constant that ensures the sum of all probabilities equals one. The first assumption in the Onsager theory consists in postulating that the entropy variation is a bilinear expression of fluctuations. Prigogine generalized Eq.~(\ref{Mm1}), which applies only to adiabatic or isothermal transformations, by introducing the entropy production due to fluctuations. Denoting by $\xi_i$ ($i=1\cdots m$) the $m$ deviations of the thermodynamic quantities from their equilibrium value, Prigogine proposed that the probability distribution of finding a state in which the values $\xi_i$ lie between $\xi_i$ and $\xi_i+d\xi_i$ is given by \cite{prigogine2}
\begin{equation}\label{Mm2}
\mathcal{W}=W_0\exp[\Delta_{\rm I}  S/k_B]\qquad\quad
{\rm where}\qquad \Delta_{\rm I}   S=\int_E^F d_{\rm I} s\quad  {\rm ;}\quad \frac{d_{\rm I}  s}{dt}\equiv\int_\Omega\sigma dv
\end{equation}
\noindent $dv$ is a (spatial) volume element of the system, and the integration is over the entire space $\Omega$ occupied by the system in question. $E$ and $F$ indicate the equilibrium state and the state to which a fluctuation has driven the system, respectively. Note that this probability distribution remains unaltered for flux-force transformations leaving invariant the entropy production. The expressions for the entropy production, in the collisional regime as well as in the turbulent regime, can be easily computed. Indeed, $\sigma$ for FTU-plasma in collisional regime may directly be obtained by the relation 
\begin{equation}\label{Mm3}
g_{\mu\nu}=L_{\mu\nu}X^\mu X^\nu
\end{equation}
\noindent From Eq.~(\ref{Mm3}) we get
\begin{equation}
\sigma=\exp(\phi) L_{\mu\nu}X^\mu X^\nu
\end{equation}
\noindent with the conformal scalar $\phi$ solution of Eq.~(\ref{TPT15}) and satisfying the boundary condition illustrated in Fig.~\ref{fig_BC1}. Fig.~\ref{fig_entropy_coll} plots the entropy production of the FTU-plasma in collisional regime in coordinates ($x,z$). 
%%%%%%%%%%%%%%%%%%%%%%%%%%%%%%%%%%%%%%%%%%
%%%%%%%%%%%%%%%%%%%%%%%%%%%%%%%%%%%%%%%%%%
\noindent The entropy production in the turbulent region can easily be obtained by recalling the relation
\begin{equation}
\sigma=\Bigl(\sigma_L{\sigma_{Onsager}}\Bigr)^{1/2}\nonumber
\end{equation}
\noindent with $\sigma_L$ solution of Eq.~(\ref{T7}). The entropy production of the FTU-plasma in a turbulent regime is shown in Fig.~\ref{fig_entropy_turb}.
%%%%%%%%%%%%%%%%%%%%%%%%%%%%%%%%%%%%%%%%%%
%%%%%%%%%%%%%%%%%%%%%%%%%%%%%%%%%%%%%%%%%%
\noindent Knowing the expression of entropic production, we find the probability of a fluctuation for plasma in the collisional and in turbulent regimes by using Eq.~(\ref{Mm2}) . The determination of the probability of a fluctuation allows for performing investigations at a microscopic level. 

\subsection{Evaluation of the Bohm coefficient}\label{bohm}

\begin{figure}[h]
\centering\includegraphics[width=7.0cm]{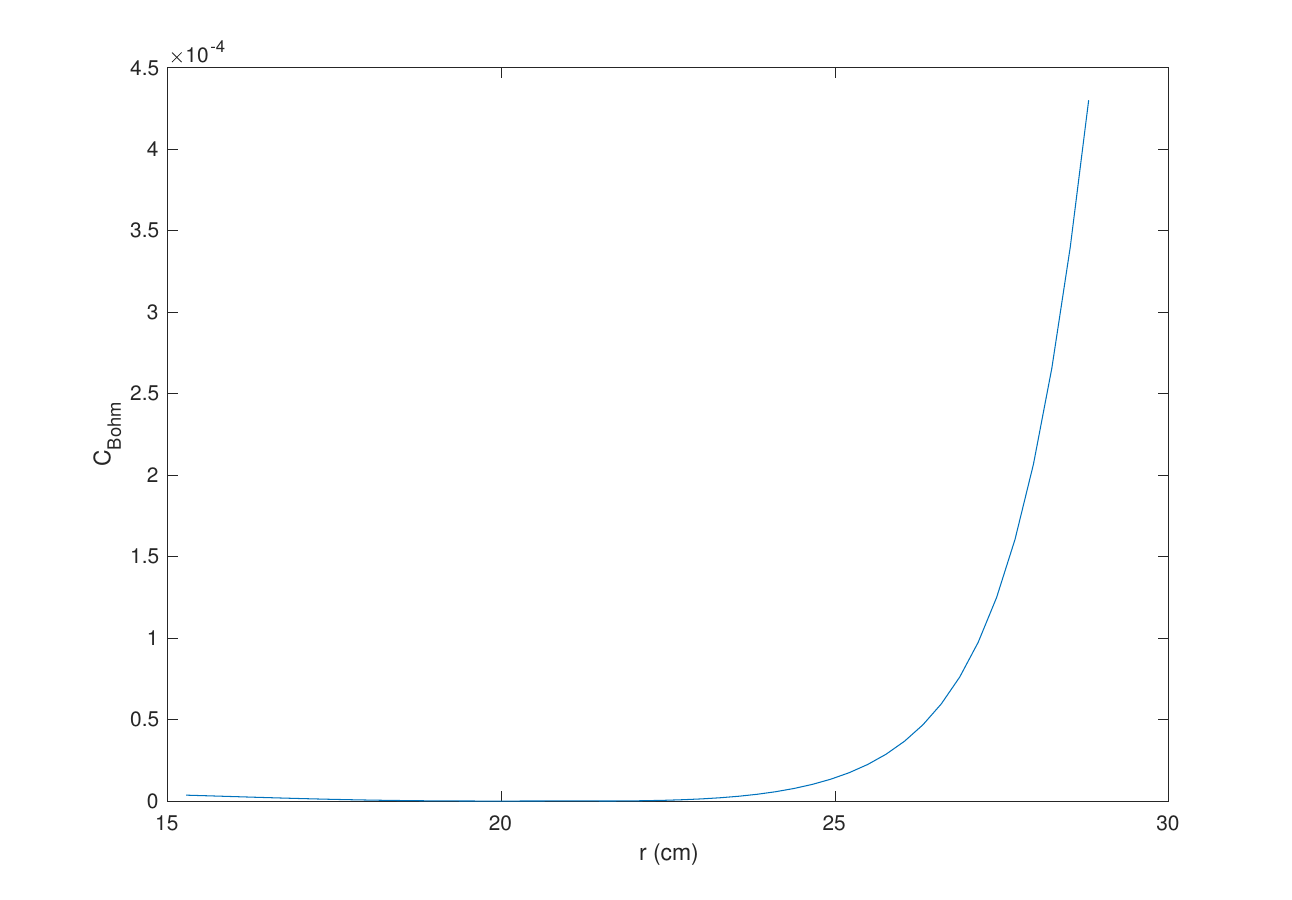}
\caption{{\bf The Bohm coefficient vs the radial coordinate $r$ of  FTU-plasmas near the edge of the Tokamak}.}
\label{fig_bohm}
\end{figure}

By taking into account Eq.~(\ref{TPT14}), Eq~(\ref{TPT2a}), and Eq.~(\ref{TPT8}), we are able to estimate the coefficient $C_{Bohm}$. Indeed, the Bohm diffusion coefficient may be brought into the form
\begin{equation}\label{Mm4}
\chi_{Bohm}=C_{Bohm}\frac{c T_\iota}{e_\iota B_0}
\end{equation}
\noindent with $c$ denoting the speed of light and $B_0$ the magnetic field at the symmetric axis of the Tokamak, respectively. Note that in Eq.~(\ref{Mm4}) we used the {\it Gauss c.g.s. unit system} and Temperature is measured in $Kev$. The expression for the diffusion coefficient in the Pfirsch-Schl$\ddot{{\rm u}}$ter (PS) transport regime is derived by kinetic theory (see, for instance, \cite{balescu2} \footnote{Note that in \cite{balescu2} the thermodynamic forces are normalised. So, in \cite{balescu2} the diffusion coefficient is defined as $L_{11}^{PS}=n_\iota \chi^{PS}$ with $n_\iota$ denoting the particle number of the species$\ \iota$.}):
\begin{equation}\label{Mm5}
\chi^{PS}=\frac{1}{2}\frac{\rho_{\iota 0}^2}{\tau_{\iota 0}}{\mathcal F}^2({\mathcal G}-1)
\end{equation}
\noindent where the {\it geometrical factor} ${\mathcal F}$ and the {\it averaged function of the magnetic field} ${\mathcal G}$ are defined as
\begin{equation}\label{Mm6}
{\mathcal F}=\frac{Rq}{r}\quad ;\quad {\mathcal G}=\Big<\frac{\beta_0^2}{B^2}\Big>\qquad{\rm with}\qquad \beta_0=(<B^2>)^{1/2}
\end{equation}
\noindent In Eq.~(\ref{Mm5}), $\rho_{\iota 0}$ and $\tau_{\iota 0}$ denote the {\it thermal Larmor radius} and {\it thermal collision time} evaluated at the averaged magnetic field $\beta_0$, respectively. $R$ and $q$ are the major radius of the Tokamak and the {\it safety factor}, respectively. By noticing that \cite{balescu2} 
\begin{equation}\label{Mm7}
\frac{\rho_{\iota 0}^2}{\tau_{\iota 0}}=2\frac{cT_\iota}{e_\iota\beta_0 x_\iota}\simeq 2\frac{cT_\iota}{e_\iota B_0 x_\iota}\qquad{\rm with}\qquad x_{\iota 0}\equiv \Omega_{\iota 0}\tau_{\iota 0}
\end{equation}
\noindent where we have taken into account that in the standard model we have $\beta_0=B_0+O(\eta^2)$ with $\eta$ denoting the inverse of the aspect ratio (i.e., $\eta=a/R$)\cite{balescu2}, we get
\begin{equation}\label{Mm8}
\chi^{PS}=\frac{cT_\iota}{e_\iota B_0 x_{\iota 0}}{\mathcal F}^2({\mathcal G}-1)
\end{equation}
\noindent In Eq.~(\ref{Mm7}) $\Omega_{\iota 0}$ is the {\it thermal Larmor frequency evaluated at the averaged field} $\beta_0$. From Eq.~(\ref{TPT14}), we have
\begin{equation}\label{Mm8}
\chi_{Turb}=\frac{c T_\iota}{e_\iota B_0 x_{\iota 0}}{\mathcal F}^2({\mathcal G}-1)<\exp(\phi)>
\end{equation}
\noindent with the scalar field $\phi$ solution of Eq.~(\ref{TPT15}) in the turbulent regime (see Section~\ref{turbulent_edge}). By comparing Eq.~(\ref{Mm4}) with Eq.~(\ref{Mm8}), we finally get
\begin{equation}\label{Mm9}
C_{Bohm}=\frac{{\mathcal F}^2}{x_{\iota 0}}({\mathcal G}-1)<\exp(\phi)>
\end{equation}
\noindent  Fig.~\ref{fig_bohm} shows the behaviour of the Bohm coefficient versus the minor radius of FTU close to the edge of the Tokamak.
%%%%%%%%%%%%%%%%%%%%%%%%%%%%%%%%%%%%%%%%%%
%%%%%%%%%%%%%%%%%%%%%%%%%%%%%%%%%%%%%%%%%%

\subsection {Evaluation of the gyro-Bohm coefficient}\label{gyrobohm}

\begin{figure}[h]
\centering\includegraphics[width=7.0cm]{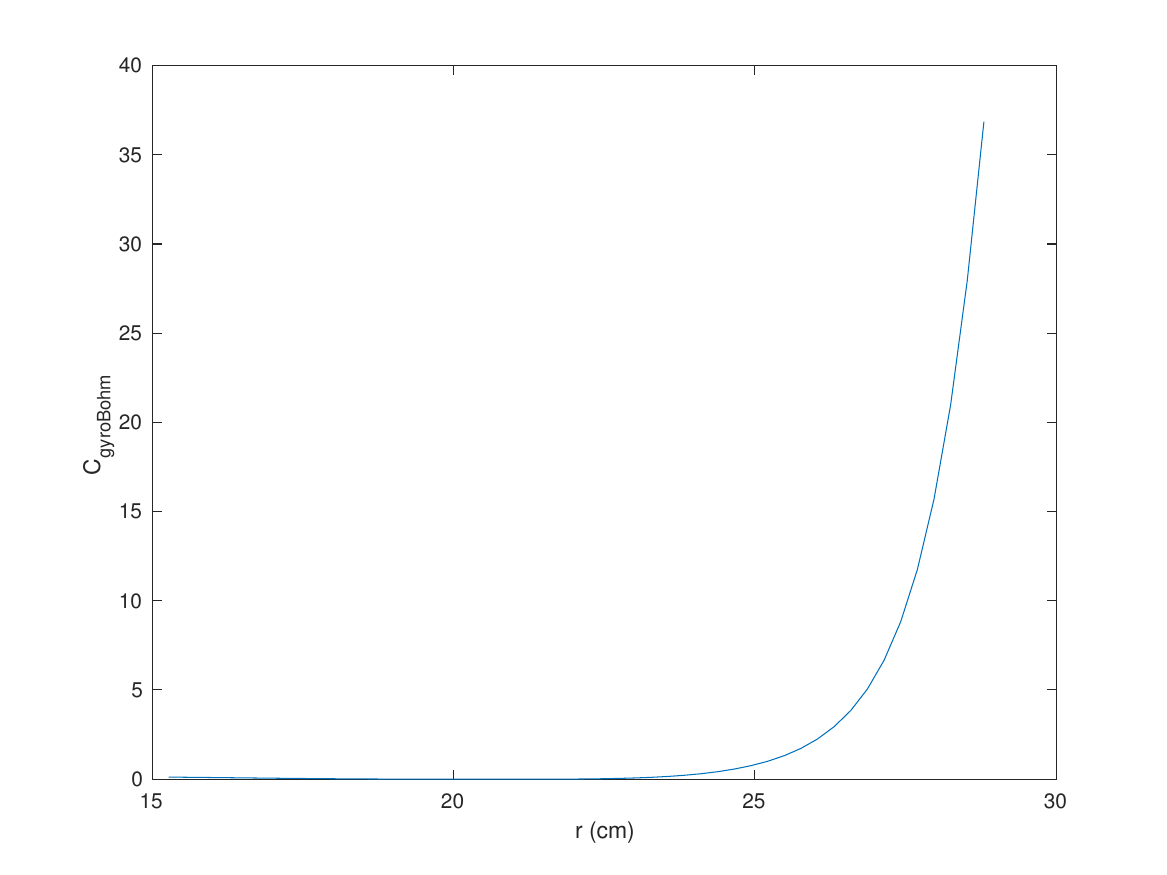}
\caption{{\bf The gyro-Bohm coefficient vs the radial coordinate $r$ of FTU-plasmas close to the edge of the Tokamak}.}
\label{fig_gyrobohm}
\end{figure}

\noindent To obtain the coefficient $C_{gyroBohm}$ we first recall the expression of the thermal Larmor radius evaluated at the averaged magnetic field $\beta_0$ \cite{balescu2}:
\begin{equation}\label{Mm10}
\frac{\rho_\iota}{a}=\frac{2^{1/2}c\ m_\iota^{1/2}T_\iota^{1/2}}{ae_\iota\beta_0}= \frac{v_{T\iota}}{a\omega_{\iota 0}}=\frac{1}{\tau_{T\iota}\omega_{\iota 0}}
\end{equation}
\noindent with $v_{T\iota}=(2T_\iota/m_\iota)^{1/2}$ and $\tau_{T\iota}\equiv a/v_{T\iota}$, respectively. By taking into account Eq.~(\ref{Mm8}) and Eq,~(\ref{Mm10}), from a comparison with Eq.~(\ref{TPT9}) we get 
\begin{equation}\label{Mm11}
C_{gyroBohm}=\frac{\tau_{T\iota}}{\tau_{\iota 0}}{\mathcal F}^2({\mathcal G}-1)<\exp(\phi)>
\end{equation}
\noindent Fig.~\ref{fig_gyrobohm} illustrates the behavior of the gyro-Bohm coefficient versus the minor radius of FTU near the edge of the Tokamak. To investigate the diffusion coefficient at the microscopic level we make reference to the Brownian motion, which establishes \cite{balescu2}
\begin{equation}\label{Mm12}
\chi_{Turb}= \frac{<(\Delta r)^2>}{\tau_{char.}}
\end{equation}
\noindent where the characteristic time $\tau_{char.}$ is the time between two random jumps of length $\Delta r$. The collisions make the particles jump on other magnetic lines. We may interpret this effect as a random walk whose time between two jumps is the thermal collision time evaluated at the averaged magnetic field $\beta_0$ (i.e., $\tau_{char.}=\tau_{\iota 0}$). In the so-called {\it hydrodynamical regime}, the relaxation time is much shorter than the hydrodynamical times: $\tau_{T\iota}/\tau_{\iota 0}\gg 1$. Hence, the gyro-Bohm coefficient is much larger than both the neoclassical coefficient and the Bohm coefficient.
%%%%%%%%%%%%%%%%%%%%%%%%%%%%%%%%%%%%%%%%%%
%%%%%%%%%%%%%%%%%%%%%%%%%%%%%%%%%%%%%%%%%%
\noindent To the best of our knowledge, this is the first time that the values for the Bohm and gyro-Bohm coefficients have been predicted theoretically. Further calculations at the microscopic level will be the subject of future work. 

\section{Perspectives}\label{perp}
The agreement between theory and experiments for FTU plasma encourages us to analyze more complex situations such as DTT and ITER. However, as previously mentioned, it should be borne in mind that in these cases the plasma is subjected to four thermodynamic forces i.e., the {\it electronic temperature gradient}, the {\it ionic temperature gradient}, the {\it pressure gradient}, and the {\it thermodynamic force linked to the external auxiliary power}. A curved space is characterized by the Riemannian tensor $R_{\lambda\mu\nu\kappa}$. The number $C_N$ of algebraically independent component of $R_{\lambda\mu\nu\kappa}$ for an $N-$dimensional space is $C_N=N^2(N^2-1)/12$ (see, for instance, \cite{weinberg}). Hence, 
\vskip0.2truecm
\noindent - In one-dimension we have $C_1 = 0$ independent components;

\noindent - In two-dimensions we have $C_2 = 1$ independent components.
\vskip0.2truecm
\noindent For a 4-dimensional thermodynamic space the Riemannian tensor possesses $C_4 = 20$ independent components ! Furthermore, the metric tensor is no longer conformal to Onsager’s matrix and must satisfy the tensorial differential equations reported in ref.~\cite{sonninochaos}. For the 4-dimensional case, the Bohm and gyro-Bohm coefficients correspond to the first component of the metric tensor $g_{11}$ but, unlike the two-dimensional case, the {\it amplification factor} differs from one component of the metric to another. Anyhow, the present work is relevant for the future generation of Tokamaks. Indeed, the results found in this work suggest that {\it the isotropic assumption relating the transport coefficients may be applied also for dealing with the 4-dimensional case}. Thus, the next step will be to apply our approach to DTT and ITER. To accomplish this task, we have:
\vskip0.2truecm
\noindent {\bf 1)} {\it to assume the validity of the isotropic hypothesis for transport coefficients};

\noindent {\bf 2)} {\it to solve, most likely numerically, the (much) more complex {\it tensorial} transport equations reported in \cite{sonninochaos}, with the appropriate boundary conditions, for DTT and ITER};

\noindent and, once obtained the expression of the component $g_{11}$, 

\noindent {\bf 3)} {\it to derive the expression for the gyro-Bohm coefficient}.
\vskip0.2truecm
\noindent This will be the subject of future work.

\section{Conclusions}\label{conclusions}
In recent work, in the framework of the thermodynamics of irreversible processes, we derived the explicit nonlinear equations, with the appropriate boundary conditions, that have to be satisfied by the transport coefficients for systems out of the Onsager region. It is worth mentioning that the model was built only by taking into account the theorems valid for systems out-of-equilibrium. No other {\it ingredients} have been added. In \cite{sonninochaos} we have shown a fairly good agreement between the theoretical predictions of our model and the experimental data for collisional  FTU-plasma. Discrepancies were, however, observed toward the edge of the Tokamak, where transport is mainly dominated by turbulence. This work is focused on the analysis of Tokamak-plasmas in turbulent regimes. Since the above equations for the transport coefficients have been derived without neglecting any term present in the dynamic equations (i.e., the energy, mass, and momentum balance equations), it is quite natural to propose this equation as a good candidate also for describing transport in two-dimensional turbulent systems. We have computed the electronic mass and heat losses for FTU-plasmas in turbulent regimes. To accomplish this task, we have adopted the following assumption: {\it when plasma is in a turbulent regime, transport coefficients are isotropic in the space of the thermodynamic forces}. This means that the transport coefficients depend only on the bilinear expression of $\sigma_L$. We showed the agreement between the theoretical predictions and the experimental data for FTU-plasmas even on the edge of the Tokamak where transport is mainly dominated by turbulence. From the microscopic point of view, we showed that field theory, combined with the theorems established by the thermodynamics of irreversible processes, is able to evaluate the Bohm and gyro-Bohm coefficients. To the best of our knowledge, this is the first time that such coefficients have been determined analytically. For easy reference, we briefly summarise the main results obtained in this work:
\vskip 0.2truecm
\noindent {\bf 1)} We have analyzed the transport processes of FTU-plasmas in turbulent regions. For this we have solved, for the first time, the differential equations to be satisfied by the transport coefficients;

\noindent {\bf 2)} We have introduced  the {\it isotropic assumption} for Tokamak-plasmas in turbulent conditions: {\it the transport coefficients have the same intensity regardless of the direction in space of the thermodynamic forces};

\noindent {\bf 3)} We showed the agreement between the theoretical predictions and experimental data for FTU-plasmas. Note that it is the first time that such an agreement has been reached. Before this work the discrepancy between theory and experiments was several orders of magnitude;

\noindent {\bf 4)} We have found, for the first time, the analytical expression for the Bohm and gyro-Bohm turbulent coefficients. Notice that, before this work, these coefficients were evaluated only numerically or through a fitting of experimental data.

\noindent {\bf 5)} For the first time, we have derived the expression of Prigogine's probability distribution function for the fluctuations of thermodynamic quantities.
\vskip 0.2truecm
\noindent The ultimate aim of this series of works is to apply our approach to more complex Tokamak-plasmas such as the Divertor Tokamak Test facility (DTT)-plasma \cite{DTT} or the International Thermonuclear Experimental Reactor (ITER)-plasmas \cite{ITER}.

\noindent We conclude with some comments about the validity of Eq.~(\ref{I1}). It is known that the most general flux-force transport relations take the form \cite{balescu3}
\begin{equation}\label{r1}
J_\mu({\bf r},t)=\int_{\Omega}d{\bf r}'\int_0^t dt'{\mathcal G}_{\mu\nu}[X({\bf r}',t')]X^{'\!\nu}({\bf r}-{\bf r}',t-t')
\end{equation}
\noindent with $\Omega$ denoting the volume occupied by the system. The space-time dependent coefficients ${\mathcal G}_{\mu\nu}$ are called \textit{nonlocal transport coefficients}: they should not be confused with coefficients $\tau_{\mu\nu}$ (they do not have the same dimension). The nonlocal and non-Markovian Eq.~(\ref{r1}) expresses the fact that the flux at a given point $({\bf r},{\rm t})$ could be influenced by the values of the forces in its spatial environment and by its history. Whenever the spatial and temporal ranges of influence are sufficiently small, the delocalization and the retardation of the forces can be neglected under the integral,
\begin{align}\label{r2}
&{\mathcal G}_{\mu\nu}[X({\bf r}',t')]X^{'\!\nu}({\bf r}-{\bf r}',t-t')\\
&=2\varpi_{\mu\nu}[X({\bf r},t)]X^\nu({\bf r},t)\delta({\bf r}-{\bf r}')\delta(t-t')\nonumber
\end{align}
\noindent with $\delta$ denoting Dirac's delta function. In this case, the transport equations reduce to
\begin{equation}\label{r3}
J_\mu({\bf r},t)=\tau_{\mu\nu}[X({\bf r},t)]X^\nu({\bf r},t)
\end{equation}
\noindent In the vast majority of cases studied at present in transport theory, it is assumed that the transport equations are of the form of Eq.~(\ref{r3}). However, equations of the form~(\ref{r1}) may be met when we deal with anomalous transport processes such as, for example, transport in turbulent tokamak plasmas - see, for example, \cite{balescu1}. Hence, Eqs~(\ref{r2}) establish, in some sort, the limit of validity of Eq.~(\ref{I1}) and, in this case, the fluxes should be evaluated by using Eq.~(\ref{r1}). Nonetheless, we would like to stress the following. Hydrodynamic turbulence is normally studied through the Navier-Stokes equations, supported by the conservation equations for the mass and energy (the so-called \textit{mass-energy balance equations}). The set of hydrodynamic equations are closed through relations of the form~(\ref{r3}) where, for Newtonian fluids, $\tau_{\mu\nu}$ depends only on the thermodynamical quantities, and not on their gradients. The experimental data are in excellent agreement with the numerical simulations - see, for example, \cite{kollmann}. For non-Newtonian fluids, turbulence is still analyzed by closing the balance equations with equations of the form~(\ref{r3}) where the viscosity coefficients depend not only on the thermodynamic quantities but also on their gradients - see, for example, \cite{lumley}. Also, in this case, the experimental data are in excellent agreement with the numerical simulations. Even transport phenomena in Tokamak-plasmas in the weak-collisional regime are analyzed by closing the balance equations with equations of the type~(\ref{r3}) - see, for example, \cite{balescu2}. This is for saying that Eqs~(\ref{r3}) is {\it very robust} equations and their validity goes well beyond the collisional, or the weak-collisional, regime. This case is very similar to what happens for the Onsager reciprocity relations: even if, according to the non-equilibrium statistical physics and the kinetic theory, these relations should have been valid only in the vicinity of the thermodynamic equilibrium in reality their validity goes well beyond the thermodynamic equilibrium, up to be valid even in turbulent hydrodynamic regimes. 

\noindent In conclusion, before further complicating the mathematical formalism, it is the author's opinion that it is still worth analyzing the turbulence in Tokamak plasmas by closing the balance equations with local equations of the type~(\ref{r3}) and comparing {\it a posteriori} the theoretical predictions with experimental data.

\section*{Data Availability Statement}

\noindent The data that support the findings of this study are available from the corresponding author upon reasonable request.

\section*{Acknowledgements}

\noindent We would like to pay tribute to our colleague and friend Prof. Enrique Tirapegui, co-author of several manuscripts of this series of works. We are grateful to Dr M. Marinucci from the ENEA - Frascati (Rome-Italy) for having provided the experimental data for FTU-plasmas. We are also indebted to Dr F. Zonca for his constant encouragement and useful scientific suggestions.

%%%%%%%%%%%%%%%%%%%%%BIBLIOGRAPHY%%%%%%%%%%%%%%%%%%%%%%
%\section*{References}

%%%%%%%%%%%%%%%%%%%%%END_BIBLIOGRAPHY%%%%%%%%%%%%%%%%%%%
\end{document}